# *Topological classification of Chern-type insulators with the photonic Green function*

*Mário G. Silveirinha*[*]

[(1)] *University of Lisbon–Instituto Superior Técnico and Instituto de Telecomunicações, Avenida Rovisco Pais, 1, 1049-001 Lisboa, Portugal, mario.silveirinha@co.it.pt*

**Abstract**

The Chern topological numbers of a material system are traditionally written in terms of the Berry curvature which depends explicitly on the material band structure and on the Bloch eigenwaves. Here, we demonstrate that it is possible to calculate the gap Chern numbers of a photonic platform without having any detailed knowledge of its band structure, relying simply on the system photonic Green function. It is shown that the gap Chern number is given by an integral of the photonic Green function along a line of the complex frequency plane parallel to the imaginary axis. Our theory applies to arbitrary frequency dispersive fully three-dimensional photonic crystals, as well as to the case of electromagnetic continua with no intrinsic periodicity.

---

[*] To whom correspondence should be addressed: E-mail: *mario.silveirinha@co.it.pt*

# I. Introduction

The last decades have witnessed a flurry of interest in topological matter. Starting with the pioneering discovery of a quantized Hall conductivity [1], it has been shown that the topology of the electronic band structure may influence decisively the electronic transport [2, 3, 4, 5]. Remarkably, insulating materials can be organized into different classes of equivalence, being each class characterized by a topological invariant (an integer), which is absolutely insensitive to weak modifications of the microstructure.

Furthermore, it has been shown that topological ideas can also be extended to photonics [6, 7], and in particular several works have highlighted that optical systems may have a topological nature [8, 9, 10, 11, 12, 13]. Rather extraordinarily, an interface of two topologically distinct photonic insulators (i.e., structures with a complete electromagnetic band gap) supports unidirectional scattering-immune edge states [10]. Hence, topological materials may enable a more efficient light transport weakly sensitive to imperfections, defects and deformations of the propagation path.

In this work, we focus on systems with a broken time-reversal symmetry. Usually, the topological classification of such materials is done using Chern invariants. The foundations of the theory of Chern-type photonic insulators were laid by Raghu and Haldane [2, 3], and more recently were extended to general bianisotropic platforms and to electromagnetic continua, i.e., systems with no intrinsic periodicity [11]. Similar to electronics, the Chern invariant is determined by the photonic band structure and by the Bloch eigenstates. Specifically, the Chern number is written in terms of a rather abstract gauge-dependent Berry potential, which relies explicitly on the normal modes of the system. A nontrivial Chern number indicates the impossibility of finding a globally defined smooth gauge of eigenfunctions. From a computational point of view, the problem of calculating the Chern invariants is rather complex: it generally requires finding the photonic band structure and all the Bloch states in the Brioullin zone. The problem is specially challenging in the case of periodic systems (nonreciprocal photonic crystals).

Remarkably, in this article it is shown that there is an unsuspected link between the system photonic Green function and the Chern invariants. We prove that the gap Chern numbers (i.e., the sum of the individual Chern numbers for the bands lying below a certain band gap) can be written in terms of the photonic Green function without any detailed knowledge of the band structure or of the Bloch eigenstates. The gap Chern number is given by an integral of the photonic Green function over a line a parallel to the imaginary axis in the complex frequency plane.

It should be mentioned that a somewhat related result is known to hold in the electronic case [14, 15], but its generalization to photonics is not evident due to the complications stemming from the dispersive response of photonic materials. Furthermore, the structure of the formula reported in Ref. [14] is totally different from the result derived here. Indeed, our formula is not an extension of the electronic case and its derivation appears to be unrelated to that of the electronic case.

The article is organized as follows: Section II presents an overview of the theory of Ref. [11], which provides for a Hermitian formulation of the electrodynamics of dispersive media. In Sect. III, we develop a theory of modal expansions in dispersive photonic platforms and use it to obtain a decomposition of the system Green function in terms of eigenmodes. In Sect. IV, a few useful formulas related to the notions of Berry potential, Berry curvature, and Chern number are presented. The main result of the article is demonstrated in Sect. V and the link between the gap Chern number and the photonic Green function is established. In Sect. VI, we apply the developed concepts to the particular case of electromagnetic continua. The relation between the Chern number and the Green function is numerically demonstrated for a magnetized electric plasma. A brief summary of the main findings is given in Sect. VII.

## II. Hermitian formulation of the electrodynamics of dispersive media

In what follows, we present a brief overview of the theory of Ref. [11], which enables characterizing the time evolution of arbitrary bianisotropic and possibly nonreciprocal and lossless systems with a Schrödinger-type formalism. Furthermore, we extend the theory to the case of inhomogeneous (in space) photonic platforms (e.g., a photonic crystal), and hence allow the material parameters to vary in space arbitrarily.

The electromagnetic response of a generic linear system can be described by a 6×6 material matrix $\mathbf{M}$, which relates the frequency domain electromagnetic fields as follows:

$$\mathbf{g}(\mathbf{r},\omega) = \mathbf{M}(\mathbf{r},\omega) \cdot \mathbf{f}(\mathbf{r},\omega), \quad \text{with} \quad \mathbf{M} = \begin{pmatrix} \varepsilon_0 \overline{\varepsilon} & \frac{1}{c}\overline{\xi} \\ \frac{1}{c}\overline{\zeta} & \mu_0 \overline{\mu} \end{pmatrix}. \tag{1}$$

Here, $\overline{\varepsilon}$ is the relative permittivity, $\overline{\mu}$ is the relative permeability and the tensors $\overline{\xi}, \overline{\zeta}$ determine the magneto-electric (bianisotropic) response [16, 17]. Except where explicitly stated otherwise (Sect. VI.B), the material response is assumed local (no spatial dispersion). We use six-vector notations with $\mathbf{f} = \begin{pmatrix} \mathbf{E} & \mathbf{H} \end{pmatrix}^T$, $\mathbf{g} = \begin{pmatrix} \mathbf{D} & \mathbf{B} \end{pmatrix}^T$, where $\mathbf{E}, \mathbf{H}$ are the electric and magnetic fields, $\mathbf{D}, \mathbf{B}$ are the electric displacement and the induction fields, and the matrix transposition operation is denoted with the superscript *T*. The electrodynamics of the system is determined by the time-domain Maxwell's equations which read:

$$\hat{N} \cdot \mathbf{f}(\mathbf{r},t) = i\left[\frac{\partial \mathbf{g}}{\partial t}(\mathbf{r},t) + \mathbf{j}(\mathbf{r},t)\right], \qquad (2)$$

with $\mathbf{j} = (\mathbf{j}_e \quad \mathbf{j}_m)^T$ the electric and magnetic current densities. The differential operator $\hat{N}$ is defined as:

$$\hat{N} = \begin{pmatrix} \mathbf{0} & i\nabla \times \mathbf{1}_{3\times 3} \\ -i\nabla \times \mathbf{1}_{3\times 3} & \mathbf{0} \end{pmatrix}, \qquad (3)$$

with $\mathbf{1}_{3\times 3}$ the identity matrix of dimension three. For simplicity, we use the same symbols to denote both the frequency domain and the time domain fields.

In Ref. [11] it was shown that the electrodynamics predicted by the time-domain Maxwell's equations is formally equivalent to the dynamics predicted by an augmented time-evolution problem described by a Hermitian operator. This result requires that $\mathbf{M}$ is a meromorphic function in the complex plane subject to the constraints $\mathbf{M}(\omega) = \mathbf{M}^*(-\omega^*)$ (reality condition), $\mathbf{M}(\mathbf{r},\omega) = \mathbf{M}^\dagger(\mathbf{r},\omega)$ for $\omega$ real-valued (lossless condition) and $\frac{\partial}{\partial \omega}[\omega \mathbf{M}(\mathbf{r},\omega)] > 0$ for $\omega$ real-valued (stored energy is nonnegative). In such a case, the material matrix has a partial-fraction expansion of the form [11]:

$$\mathbf{M}(\mathbf{r},\omega) = \mathbf{M}_\infty - \sum_\alpha \frac{\text{sgn}(\omega_{p,\alpha})}{\omega - \omega_{p,\alpha}} \mathbf{A}_\alpha^2. \qquad (4)$$

Here, $\mathrm{sgn} = \pm$ is the sign function, $\mathbf{M}_\infty = \lim_{\omega \to \infty} \mathbf{M}(\omega)$ gives the asymptotic high-frequency response of the material, $\omega_{p,\alpha}$ are the (real-valued) poles of $\mathbf{M}$, and $\mathbf{A}_\alpha = \left[ -\mathrm{sgn}(\omega_{p,\alpha})(\mathrm{Res}\mathbf{M})_\alpha \right]^{1/2}$, with $(\mathrm{Res}\mathbf{M})_\alpha$ the residue of the pole $\omega_{p,\alpha}$. The matrix $\mathbf{A}_\alpha$ is a positive (semi-)definite Hermitian matrix ($\mathbf{A}_\alpha \geq 0$). Evidently, in a inhomogeneous system, $\omega_{p,\alpha}$, $\mathbf{A}_\alpha$, and $\mathbf{M}_\infty$ are functions of $\mathbf{r}$. The sum in Eq. (4) is over all poles, including the negative frequency poles.

The augmented generalized problem describes the time evolution of a state vector of the form, $\mathbf{Q} = \begin{pmatrix} \mathbf{f} & \mathbf{Q}^{(1)} & \ldots & \mathbf{Q}^{(\alpha)} & \ldots \end{pmatrix}^T$. Each component of $\mathbf{Q}$ is a six-component vector, and the number of components depends on the number of poles of the material matrix [11]. The first component of the state vector, $\mathbf{f}$, gives the electromagnetic fields. The remaining components, $\mathbf{Q}^{(\alpha)}$, describe the internal degrees of freedom of the material response [8, 11, 18-20].

The time-evolution of the state vector is determined by a differential equation, $\hat{L} \cdot \mathbf{Q}(\mathbf{r},t) = i \frac{\partial}{\partial t} \mathbf{M}_g \cdot \mathbf{Q}(\mathbf{r},t) + i \mathbf{j}_g(\mathbf{r},t)$, which may be spelled out as [11]:

$$\underbrace{\begin{pmatrix} \hat{N} + \sum_\alpha \mathrm{sgn}(\omega_{p,\alpha}) \mathbf{A}_\alpha^2 & |\omega_{p,1}|^{1/2} \mathbf{A}_1 & |\omega_{p,2}|^{1/2} \mathbf{A}_2 & \ldots \\ |\omega_{p,1}|^{1/2} \mathbf{A}_1 & \omega_{p,1}\mathbf{1} & 0 & \ldots \\ |\omega_{p,2}|^{1/2} \mathbf{A}_2 & 0 & \omega_{p,2}\mathbf{1} & \ldots \\ \ldots & \ldots & \ldots & \ldots \end{pmatrix}}_{\hat{L}} \cdot \underbrace{\begin{pmatrix} \mathbf{f} \\ \mathbf{Q}^{(1)} \\ \mathbf{Q}^{(2)} \\ \ldots \end{pmatrix}}_{\mathbf{Q}} = i\frac{\partial}{\partial t} \underbrace{\begin{pmatrix} \mathbf{M}_\infty & 0 & 0 & \ldots \\ 0 & 1 & 0 & \ldots \\ 0 & 0 & 1 & \ldots \\ \ldots & \ldots & \ldots & \ldots \end{pmatrix}}_{\mathbf{M}_g} \cdot \begin{pmatrix} \mathbf{f} \\ \mathbf{Q}^{(1)} \\ \mathbf{Q}^{(2)} \\ \ldots \end{pmatrix} + i \underbrace{\begin{pmatrix} \mathbf{j} \\ 0 \\ 0 \\ \ldots \end{pmatrix}}_{\mathbf{j}_g}$$

(5)

with $\mathbf{1} \equiv \mathbf{1}_{6 \times 6}$ the identity matrix. The result (5) was originally derived for an electromagnetic continuum, but a straightforward modification of the original derivation shows that it also applies when the material parameters vary in space in an arbitrary way [18], for example, in case of a photonic crystal. For inhomogeneous systems, both $\hat{L}$ (which is a differential operator) and $\mathbf{M}_g$ depend explicitly on the spatial coordinates. Furthermore, since both $\hat{L}$ and $\mathbf{M}_g$ are Hermitian with respect to the canonical inner product, the operator $\hat{H}_g = \mathbf{M}_g^{-1} \cdot \hat{L}$ is Hermitian with respect to the weighted inner product:

$$\langle \mathbf{Q}_B | \mathbf{Q}_A \rangle \equiv \int_V \frac{1}{2} \mathbf{Q}_B^* \cdot \mathbf{M}_g(\mathbf{r}) \cdot \mathbf{Q}_A d^3\mathbf{r}. \tag{6}$$

Here, $V$ is the volume of interest and it is implicit that the boundary conditions (e.g., periodic boundary conditions) ensure that $\hat{H}_g$ is indeed Hermitian. In particular, without an external excitation ($\mathbf{j}_g = 0$), the augmented system (5) reduces to $\hat{H}_g \cdot \mathbf{Q} = i \frac{\partial}{\partial t} \mathbf{Q}$, which is formally equivalent to the Schrödinger equation with $\hbar = 1$.

For future reference, we note that the frequency domain $\mathbf{Q}^{(\alpha)}$ component of the state vector is related to the frequency domain electromagnetic fields as:

$$\mathbf{Q}^{(\alpha)}(\mathbf{r},\omega) = \frac{|\omega_{p,\alpha}|^{1/2}}{(\omega - \omega_{p,\alpha})} \mathbf{A}_\alpha \cdot \mathbf{f}(\mathbf{r},\omega). \tag{7}$$

This formula holds even when the electromagnetic excitation $\mathbf{j}$ is nontrivial.

## III. The Photonic Green function

Given some photonic system, we introduce a frequency domain Green function $\overline{\mathbf{G}}(\mathbf{r},\mathbf{r}',\omega)$ defined as the solution of [21, 22]:

$$\hat{N} \cdot \overline{\mathbf{G}}(\mathbf{r},\mathbf{r}',\omega) = \omega \mathbf{M}(\mathbf{r},\omega) \cdot \overline{\mathbf{G}}(\mathbf{r},\mathbf{r}',\omega) + i\mathbf{1}\delta(\mathbf{r}-\mathbf{r}'). \tag{8}$$

with $\mathbf{1} \equiv \mathbf{1}_{6\times 6}$. The Green function is a 6×6 tensor and may be decomposed into electric and magnetic terms as follows:

$$\overline{\mathbf{G}} = \begin{pmatrix} \mathbf{G}_{EE} & \mathbf{G}_{EM} \\ \mathbf{G}_{ME} & \mathbf{G}_{MM} \end{pmatrix}. \tag{9}$$

When the magnetic response is trivial ($\overline{\mu} = \mathbf{1}_{3\times 3}$, $\overline{\xi} = \overline{\zeta} = 0$) the electric component of the Green function ($\mathbf{G}_{EE}$) can be written as $\mathbf{G}_{EE} = i\omega\mu_0 \mathcal{G}$ where $\mathcal{G}$ is the standard electric Green function that satisfies $\nabla \times \nabla \times \mathcal{G} - \frac{\omega^2}{c^2}\overline{\varepsilon} \cdot \mathcal{G} = \mathbf{1}_{3\times 3}\delta(\mathbf{r}-\mathbf{r}')$. In the following, we obtain a formal expansion for the Green function in terms of the electromagnetic modes of a dispersive material system. The following derivation extends the results of Ref. [21], which assume that the observation point lies in a free-space region.

## A. Scalar product of two time-harmonic fields

As previously mentioned, it is possible to introduce in a natural way a weighted scalar product [Eq. (6)] in the augmented space wherein the state vector $\mathbf{Q}$ is defined. Interestingly, it was shown in Ref. [11] that in some conditions the inner product of two state vectors may be simply expressed in terms of the corresponding electromagnetic components. Specifically, let $\mathbf{Q}_A(\mathbf{r},t) = \tilde{\mathbf{Q}}_A(\mathbf{r})e^{-i\omega_A t}$ and $\mathbf{Q}_B(\mathbf{r},t) = \tilde{\mathbf{Q}}_B(\mathbf{r})e^{-i\omega_B t}$ be two generic time-harmonic solutions of the generalized problem (5), with a time variation of the form $e^{-i\omega t}$. Furthermore, let $\mathbf{f}_A(\mathbf{r},t) = \mathbf{F}_A(\mathbf{r})e^{-i\omega_A t}$ and $\mathbf{f}_B(\mathbf{r},t) = \mathbf{F}_B(\mathbf{r})e^{-i\omega_B t}$ be the corresponding solutions of the time-harmonic Maxwell's equations, which thereby satisfy:

$$\hat{N} \cdot \mathbf{f} = \omega \mathbf{M}(\mathbf{r},\omega) \cdot \mathbf{f} + i\mathbf{j}. \tag{10}$$

Then, it is possible to show that [11, 18]

$$\langle \tilde{\mathbf{Q}}_B | \tilde{\mathbf{Q}}_A \rangle = \begin{cases} \dfrac{1}{2} \displaystyle\int_V d^3\mathbf{r}\, \mathbf{F}_B^*(\mathbf{r}) \cdot \left[ \dfrac{\omega_B \mathbf{M}(\mathbf{r},\omega_B) - \omega_A \mathbf{M}(\mathbf{r},\omega_A)}{\omega_B - \omega_A} \right] \cdot \mathbf{F}_A(\mathbf{r}), & \text{if } \omega_A \neq \omega_B \\ \dfrac{1}{2} \displaystyle\int_V d^3\mathbf{r}\, \mathbf{F}_B^*(\mathbf{r}) \cdot \dfrac{\partial}{\partial \omega}\left[\omega \mathbf{M}(\mathbf{r},\omega)\right] \cdot \mathbf{F}_A(\mathbf{r}), & \text{if } \omega_A = \omega_B \equiv \omega \end{cases}, \quad (11)$$

so that the weighted scalar product is written directly in terms of the electromagnetic field "time envelopes". It is underlined that the system parameters may vary in space. Furthermore, the scalar product of a time-harmonic solution of (5) with itself, $\langle \mathbf{Q} | \mathbf{Q} \rangle$, gives precisely the energy stored in the volume $V$.

## B. Modal expansions

Let us consider now a set of natural modes of oscillation ($\mathbf{f}_n$) of the considered photonic system, i.e. the solutions of

$$\hat{N} \cdot \mathbf{f}_n = \omega_n \mathbf{M}(\mathbf{r},\omega_n) \cdot \mathbf{f}_n, \qquad (12)$$

where $\omega_n$ are the eigenfrequencies of the problem. It is shown in Appendix A, that the eigenmodes may be chosen such that the following generalized orthogonality conditions are satisfied:

$$\frac{1}{2} \int_V d^3\mathbf{r}\, \mathbf{f}_n^*(\mathbf{r}) \cdot \frac{\partial}{\partial \omega}\left[\omega \mathbf{M}(\mathbf{r},\omega)\right]_{\omega=\omega_n} \cdot \mathbf{f}_m(\mathbf{r}) = \delta_{n,m}, \qquad \text{if } \omega_n = \omega_m. \qquad (13a)$$

$$\frac{1}{2} \int_V d^3\mathbf{r}\, \mathbf{f}_n^*(\mathbf{r}) \cdot \left[ \frac{\omega_n \mathbf{M}(\mathbf{r},\omega_n) - \omega_m \mathbf{M}(\mathbf{r},\omega_m)}{\omega_n - \omega_m} \right] \cdot \mathbf{f}_m(\mathbf{r}) = 0, \qquad \text{if } \omega_n \neq \omega_m. \qquad (13b)$$

Moreover, with this normalization the eigenmodes satisfy the completeness relation:

$$\delta(\mathbf{r}-\mathbf{r}')\mathbf{M}_\infty^{-1}(\mathbf{r}') = \frac{1}{2}\sum_n \mathbf{f}_n(\mathbf{r}) \otimes \mathbf{f}_n^*(\mathbf{r}'). \qquad (14)$$

Thus, it is always possible to expand an arbitrary electromagnetic field distribution $\mathbf{f}$ in terms of the eigenmodes $\mathbf{f}_n$, so that $\mathbf{f}(\mathbf{r}) = \sum_n \mathbf{f}_n(\mathbf{r}) c_n$. Remarkably, as further highlighted in Appendix A, in a dispersive system the coefficients of the expansion $c_n$ are *not* unique. In brief, the reason is that the modal expansions are unique only in the augmented space wherein the state vector $\mathbf{Q}$ is defined, but not in the "projection" electromagnetic space. In Appendix A, it is proven that when $\mathbf{f}(\mathbf{r})$ is some solution of the time-harmonic problem (10), it may be expanded in terms of the eigenmodes as follows:

$$\mathbf{f}(\mathbf{r}) = \sum_n \mathbf{f}_n(\mathbf{r}) c_n, \quad \text{with} \quad c_n = \frac{1}{2} \int_V d^3\mathbf{r} \, \frac{\mathbf{f}_n^* \cdot i\mathbf{j}}{\omega_n - \omega}. \tag{15}$$

### C. Modal expansion of the Green function

Applying the result (15) to the solution of (8), it is readily found that in the limit of vanishing material loss the Green function has the modal expansion:

$$\overline{\mathbf{G}}(\mathbf{r}, \mathbf{r}', \omega) = \frac{i}{2} \sum_n \frac{1}{\omega_n - \omega} \mathbf{f}_n(\mathbf{r}) \otimes \mathbf{f}_n^*(\mathbf{r}'). \tag{16}$$

It is stressed that the electromagnetic modes must be normalized as in Eq. (13). Using the completeness relation (14), it is possible to restrict the summation to "transverse" modes with $\omega_n \neq 0$, so that $\overline{\mathbf{G}} = \frac{i}{2} \sum_{\omega_n \neq 0} \frac{\omega_n}{(\omega_n - \omega)\omega} \mathbf{f}_n(\mathbf{r}) \otimes \mathbf{f}_n^*(\mathbf{r}_0) - \frac{i}{\omega} \delta(\mathbf{r} - \mathbf{r}_0) \mathbf{M}_\infty^{-1}(\mathbf{r}_0)$. Note that the summation includes both positive and negative frequency modes. It is underscored that the developed theory applies to general non-uniform, bianisotropic and possibly nonreciprocal material platforms.

## IV. Berry curvature and Chern number

Typically, the topological classification of physical systems relies on the spectrum (band structure) of some Hermitian operator [23]. Usually photonic systems are formed by dispersive materials, and hence their topological classification must be done using the eigenmodes of the generalized (homogeneous) problem (5) [8, 9, 11].

For periodic systems (as well as for electromagnetic continua [11]) the modes of the augmented problem are Bloch waves, i.e., solutions of the form $\mathbf{Q}_{n\mathbf{k}}(\mathbf{r})e^{i\mathbf{k}\cdot\mathbf{r}}$, where the periodic "spatial envelope" satisfies:

$$\hat{H}_g(\mathbf{r}, -i\nabla + \mathbf{k}) \cdot \mathbf{Q}_{n\mathbf{k}} = \omega_{n\mathbf{k}} \mathbf{Q}_{n\mathbf{k}}. \tag{17}$$

Note that the Hermitian operator $\hat{H}_g = \mathbf{M}_g^{-1} \cdot \hat{L}$ is of the form $\hat{H}_g(\mathbf{r}, -i\nabla) = \mathbf{M}_g^{-1}(\mathbf{r}) \cdot \hat{L}(\mathbf{r}, -i\nabla)$, where $\hat{L}$ depends on $-i\nabla$ only through the differential operator $\hat{N}$. Assuming the normalization $\langle \mathbf{Q}_{n\mathbf{k}} | \mathbf{Q}_{n\mathbf{k}} \rangle = 1$, the gauge dependent Berry potential is defined as:

$$\mathcal{A}_{n\mathbf{k}} = i \langle \mathbf{Q}_{n\mathbf{k}} | \partial_{\mathbf{k}} \mathbf{Q}_{n\mathbf{k}} \rangle, \tag{18}$$

where $\partial_{\mathbf{k}} = \dfrac{\partial}{\partial k_x}\hat{\mathbf{x}} + \dfrac{\partial}{\partial k_y}\hat{\mathbf{y}}$, and it is implicit that the system is closed along the *z*-direction, so that propagation is only allowed along directions parallel to the *xoy* plane (see Fig. 1). It is highlighted that the system may be fully three-dimensional: the only restrictions are the periodicity in the *xoy* plane and that the energy must flow along directions parallel to the same plane. For example, the system may be a generic periodic waveguide covered with two opaque plates (e.g., perfectly electric conducting - PEC) placed at *z*=0 and *z*=*d* (bottom and top walls, respectively). We note in passing that the Berry potential can be directly written in terms of the electromagnetic field envelope $\mathbf{f}_{n\mathbf{k}}$, defined so that $\mathbf{Q}_{n\mathbf{k}} = \begin{pmatrix} \mathbf{f}_{n\mathbf{k}} & \mathbf{Q}_{n\mathbf{k}}^{(1)} & ... \end{pmatrix}^T$ (see Refs. [8, 9, 11] for more details).

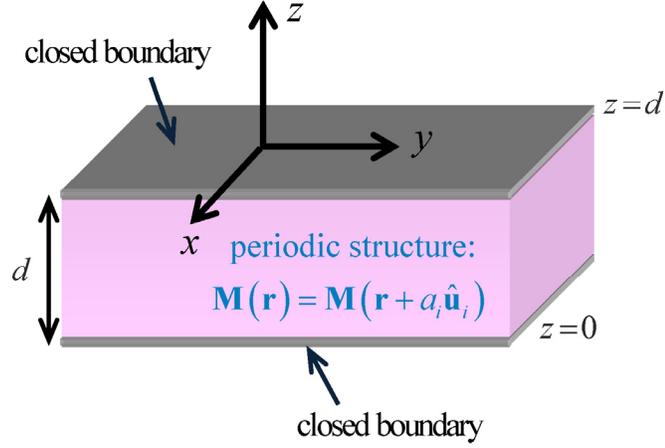

Fig. 1 Representative geometry: the system is periodic (e.g., $\mathbf{M}(x,y,z) = \mathbf{M}(x+a_x,y,z)$ and $\mathbf{M}(x,y,z) = \mathbf{M}(x,y+a_y,z)$, with $a_x, a_y$ the spatial periods along $x$ and $y$, respectively) and is electromagnetically closed so that the energy can flow only in directions parallel to the *xoy* plane.

The Berry curvature is given by $\mathcal{F}_{n\mathbf{k}} = \hat{\mathbf{z}} \cdot \nabla \times \mathcal{A}_{n\mathbf{k}}$, or equivalently [23]

$$\mathcal{F}_{n\mathbf{k}} = i\left[\langle \partial_1 \mathbf{Q}_{n\mathbf{k}} | \partial_2 \mathbf{Q}_{n\mathbf{k}} \rangle - \langle \partial_2 \mathbf{Q}_{n\mathbf{k}} | \partial_1 \mathbf{Q}_{n\mathbf{k}} \rangle \right], \tag{19}$$

where $\partial_i = \partial / \partial k_i$ ($i=1,2$) with $k_1 = k_x$ and $k_2 = k_y$. For a given complete photonic band gap, the Chern number is defined by:

$$\mathcal{C} = \frac{1}{2\pi} \iint_{B.Z.} d^2\mathbf{k} \sum_{n \in F} \mathcal{F}_{n\mathbf{k}}. \tag{20}$$

The integration region is the first Brioullin zone of the photonic crystal. The summation is over all the "filled" photonic bands (*F*) below the gap, i.e., modes with $\omega_{n\mathbf{k}} < \omega_{gap}$ (including negative frequency modes), with $\omega_{gap}$ some frequency in the band gap. The Chern number is an integer insensitive to weak deformations of the material structure, and thereby has a topological nature.

For the purposes of this study, it is convenient to write the Chern number as a discrete summation rather than as an integral. In Appendix B, it is proven that the gap Chern number can be expressed as:

$$\mathcal{C} = \frac{2\pi}{A_{tot}} \sum_{\substack{m \in E, \\ n \in F}} i \frac{1}{(\omega_n - \omega_m)^2} \left[ \langle \mathbf{Q}_n | \partial_1 \hat{H}_g | \mathbf{Q}_m \rangle \langle \mathbf{Q}_m | \partial_2 \hat{H}_g | \mathbf{Q}_n \rangle - 1 \leftrightarrow 2 \right] \quad (21)$$

where $\partial_i \hat{H}_g$ is the derivative of $\hat{H}_g(\mathbf{r}, -i\nabla + \mathbf{k})$ with respect to $k_i$ and "$1 \leftrightarrow 2$" stands for the term $\langle \mathbf{Q}_n | \partial_2 \hat{H}_g | \mathbf{Q}_m \rangle \langle \mathbf{Q}_m | \partial_1 \hat{H}_g | \mathbf{Q}_n \rangle$, i.e., the term with indices "1" and "2" interchanged. Note that a similar formula holds in the electronic case [23], with the difference that in electronics $\partial_i \hat{H}_g$ depends explicitly on the wave vector. In Eq. (21) the summation in $n$ is over the "filled" bands (modes with $\omega_n < \omega_{gap}$) and the summation in $m$ is over the "empty" ($E$) bands (modes with $\omega_n > \omega_{gap}$). Furthermore, it is implicit that the volume $V$ contains a finite number $N_x \times N_y$ of unit cells and is terminated with periodic boundaries. The transverse area of $V$ is denoted by $A_{tot}$ and the identity in (21) is strictly valid when $N_x, N_y \to \infty$. Moreover, in Eq. (21) $\{\mathbf{Q}_n\}_{n=1,2,...}$ may be understood as the "full" modes of the cavity $V$ (i.e., the solutions of $\hat{H}_g(\mathbf{r}, -i\nabla)\mathbf{Q}_n = \omega_n \mathbf{Q}_n$), rather than the "spatial envelopes". This is so because $\langle \mathbf{Q}_n | \partial_i \hat{H}_g | \mathbf{Q}_m \rangle$ is nonzero only for modes with the same wave vector, and hence it is irrelevant if the propagation factor $e^{i\mathbf{k}\cdot\mathbf{r}}$ is included or suppressed.

# V. Link between the Green function and the gap Chern number

We are finally ready to establish a link between the gap Chern number [Eq. (21)] and the photonic Green function [Eq. (16)].

## *A. Chern number as an integral in the complex frequency plane*

The first step of the proof is to show that the Chern number can be written as an integral in the complex frequency plane along the line $\text{Re}(\omega) = \omega_{\text{gap}}$ parallel to the complex imaginary axis. Here, $\omega_{\text{gap}}$ is any frequency in the relevant photonic band gap. To this end, we note that for $\omega_m \neq \omega_n$:

$$\frac{1}{(\omega-\omega_m)^2}\frac{1}{\omega-\omega_n} = \frac{1}{(\omega-\omega_m)^2}\frac{1}{\omega_m-\omega_n} + \frac{1}{(\omega_m-\omega_n)^2}\left[\frac{1}{\omega-\omega_n} - \frac{1}{\omega-\omega_m}\right] \quad (22)$$

The line $\text{Re}(\omega) = \omega_{\text{gap}}$ splits the complex frequency plane into two semi-planes (Fig. 2).

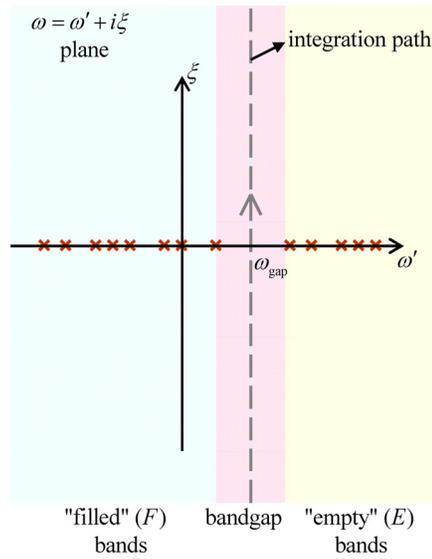

Fig. 2 Complex frequency plane showing the integration path $\text{Re}(\omega) = \omega_{\text{gap}}$ in the band gap region and the semi-planes associated with the filled bands (natural modes with $\omega_n < \omega_{\text{gap}}$) and empty bands (natural

modes with $\omega_n > \omega_{gap}$). The red crosses in the real-frequency axis illustrate possible locations for the eigenfrequencies $\omega_n$.

From Cauchy's residue theorem the integral of $\frac{1}{(\omega-\omega_m)^2}\frac{1}{\omega-\omega_n}$ over the line $\text{Re}(\omega)=\omega_{gap}$ vanishes if $\omega_m, \omega_n$ are on the same semi-plane (note that the integration contour may be closed with a semi-circle with infinite radius that does not contribute to the integral). Evidently, the same property holds when $\omega_m = \omega_n$. On the other hand, when the two poles lie in different semi-planes (e.g., if $\omega_n < \omega_{gap}$ and $\omega_m > \omega_{gap}$) the residue theorem gives:

$$\int_{\omega_{gap}-i\infty}^{\omega_{gap}+i\infty} d\omega \frac{1}{(\omega-\omega_m)^2}\frac{1}{\omega-\omega_n} = \frac{2\pi i}{(\omega_m-\omega_n)^2}\text{sgn}(\omega_{gap}-\omega_n). \tag{23}$$

Noting that in Eq. (21) the two sets $E$ and $F$ lie in different semi-planes of the complex plane (see Fig. 2), the previous formula enables us to express the Chern number as:

$$\mathcal{C} = \frac{1}{A_{tot}}\sum_{\substack{m\in E,\\ n\in F}}\int_{\omega_{gap}-i\infty}^{\omega_{gap}+i\infty} d\omega \frac{1}{(\omega-\omega_m)^2}\frac{1}{\omega-\omega_n}\left[\langle \mathbf{Q}_n|\partial_1\hat{H}_g|\mathbf{Q}_m\rangle\langle \mathbf{Q}_m|\partial_2\hat{H}_g|\mathbf{Q}_n\rangle - 1\leftrightarrow 2\right] \tag{24}$$

To proceed further, we use the fact that the Chern number can also be written as $\mathcal{C} = -\frac{2\pi}{A_{tot}}\sum_{\substack{m\in F,\\ n\in E}}(...)$, where the generic term of summation is the same as in Eq. (21) (see Appendix B). Hence, using again Eq. (23) it follows that $\mathcal{C} = \frac{1}{A_{tot}}\sum_{\substack{m\in F,\\ n\in E}}(...)$ with the summation term identical to that in Eq. (24). By averaging the two formulas one finds that $\mathcal{C} = \frac{1}{2A_{tot}}\left[\sum_{\substack{m\in E,\\ n\in F}}... + \sum_{\substack{m\in F,\\ n\in E}}...\right]$. However, because the integral of $\int_{\omega_{gap}-i\infty}^{\omega_{gap}+i\infty} d\omega \frac{1}{(\omega-\omega_m)^2}\frac{1}{\omega-\omega_n}$

vanishes when the poles are in the same semi-plane, the summations over *m* and *n* can be unconstrained, so that:

$$\mathcal{C} = \frac{1}{2A_{tot}} \sum_{m,n} \int_{\omega_{gap}-i\infty}^{\omega_{gap}+i\infty} d\omega \frac{1}{(\omega-\omega_m)^2} \frac{1}{\omega-\omega_n} \left[ \langle \mathbf{Q}_n | \partial_1 \hat{H}_g | \mathbf{Q}_m \rangle \langle \mathbf{Q}_m | \partial_2 \hat{H}_g | \mathbf{Q}_n \rangle - 1 \leftrightarrow 2 \right]. \quad (25)$$

Using Eq. (B3) and the definition of the weighted inner product [Eq. (6)] we finally obtain:

$$\mathcal{C} = \frac{1}{8A_{tot}} \sum_{m,n} \int_{\omega_{gap}-i\infty}^{\omega_{gap}+i\infty} d\omega \frac{1}{(\omega-\omega_m)^2} \frac{1}{\omega-\omega_n} \iint dV dV' \left[ \mathbf{f}_n^*(\mathbf{r}) \cdot \partial_1 \hat{N} \cdot \mathbf{f}_m(\mathbf{r}) \mathbf{f}_m^*(\mathbf{r}') \cdot \partial_2 \hat{N} \cdot \mathbf{f}_n(\mathbf{r}') - 1 \leftrightarrow 2 \right]$$
(26)

where $\partial_i \hat{N}$ is defined as in Eq. (B3) and $\mathbf{f}_n$ is the electromagnetic component of the mode $\mathbf{Q}_n$. The modes $\mathbf{f}_n$ are normalized as in Eq. (13) because $\langle \mathbf{Q}_n | \mathbf{Q}_n \rangle = 1$.

## *B. Link with the Green function*

To establish a link between the gap Chern number and the Green function we note that from the modal expansion (16) it follows that:

$$\begin{aligned}
&\text{tr}\left\{ \partial_2 \hat{N} \cdot \overline{\mathbf{G}}(\mathbf{r}, \mathbf{r}', \omega) \cdot \partial_1 \hat{N} \cdot \partial_\omega \overline{\mathbf{G}}(\mathbf{r}', \mathbf{r}, \omega) \right\} \\
&= \frac{1}{4} \sum_{m,n} \frac{-i}{\omega-\omega_n} \frac{+i}{(\omega-\omega_m)^2} \text{tr}\left[ \partial_2 \hat{N} \cdot \mathbf{f}_n(\mathbf{r}) \otimes \mathbf{f}_n^*(\mathbf{r}') \cdot \partial_1 \hat{N} \cdot \mathbf{f}_m(\mathbf{r}') \otimes \mathbf{f}_m^*(\mathbf{r}) \right], \quad (27) \\
&= \frac{1}{4} \sum_{m,n} \frac{1}{\omega-\omega_n} \frac{1}{(\omega-\omega_m)^2} \mathbf{f}_m^*(\mathbf{r}) \cdot \partial_2 \hat{N} \cdot \mathbf{f}_n(\mathbf{r}) \mathbf{f}_n^*(\mathbf{r}') \cdot \partial_1 \hat{N} \cdot \mathbf{f}_m(\mathbf{r}')
\end{aligned}$$

where "tr" stands for the trace of a tensor and $\partial_\omega \equiv \partial/\partial\omega$. Substituting this result into Eq. (26) we find that (note that the normalization of the modes is the same in Eqs. (16) and (26)):

$$\mathcal{C} = \frac{1}{2A_{tot}} \int_{\omega_{gap}-i\infty}^{\omega_{gap}+i\infty} d\omega \iint dV dV' \left[ \text{tr}\left\{ \partial_2 \hat{N} \cdot \overline{\mathbf{G}}(\mathbf{r}, \mathbf{r}', \omega) \cdot \partial_1 \hat{N} \cdot \partial_\omega \overline{\mathbf{G}}(\mathbf{r}', \mathbf{r}, \omega) \right\} - 1 \leftrightarrow 2 \right]. \quad (28)$$

Furthermore, integrating by parts in $\omega$ the term "$1 \leftrightarrow 2$" and using the cyclic property of the trace ($\mathrm{tr}\{\mathbf{A}\cdot\mathbf{B}\} = \mathrm{tr}\{\mathbf{B}\cdot\mathbf{A}\}$) one finds that the two terms inside the rectangular brackets are identical so that:

$$\mathcal{C} = \frac{1}{A_{tot}} \int_{\omega_{gap}-i\infty}^{\omega_{gap}+i\infty} d\omega \iint dV dV' \left[ \mathrm{tr}\left\{ \partial_2 \hat{N} \cdot \overline{\mathbf{G}}(\mathbf{r},\mathbf{r}',\omega) \cdot \partial_1 \hat{N} \cdot \partial_\omega \overline{\mathbf{G}}(\mathbf{r}',\mathbf{r},\omega) \right\} \right]. \tag{29}$$

This is the main result of the article. It establishes that the gap Chern number, i.e., the sum of all Chern numbers of photonic bands with $\omega_n < \omega_{gap}$ (including negative frequency bands) can be written in terms of an integral of the photonic Green function along a straight line parallel to the complex imaginary axis. The photonic Green function is the solution of Eq. (8). The identity (29) holds in the limit $A_{tot} \to \infty$, i.e., when the considered volumetric region $V$ (terminated with periodic boundaries along $x$ and $y$) becomes the entire space (unbounded photonic crystal). We remind that the definition of the Chern number assumes that $V$ is electromagnetically closed so that the energy is forced to flow along directions parallel to the *xoy* plane (Fig. 1). The simplest case is when the system is uniform along the *z*-direction and the condition $\partial/\partial z = 0$ is enforced to constrain the wave propagation to the *xoy* plane. In such a scenario, the electromagnetic problem and the associated Green function are effectively two-dimensional. This case is further discussed in Sect. V.C.

Equation (29) is manifestly gauge invariant because it is written solely in terms of the Green function and its frequency derivative in the complex frequency plane. Note that the Green function is free of singularities along the integration path because its poles (the eigenfrequencies) lie outside the band gap region. More importantly, it is highlighted that Eq. (30) is fully independent of any specific gauge, and thus it avoids the usual troubles of particular gauges not being globally defined in the wave vector space [11, 23].

It can be verified that for a reciprocal system the Green function satisfies $\overline{\mathbf{G}}(\mathbf{r},\mathbf{r}',\omega) = \boldsymbol{\sigma}_z \cdot \overline{\mathbf{G}}^T(\mathbf{r}',\mathbf{r},\omega) \cdot \boldsymbol{\sigma}_z$, $\boldsymbol{\sigma}_z = \begin{pmatrix} \mathbf{1}_{3\times 3} & \mathbf{0} \\ \mathbf{0} & -\mathbf{1}_{3\times 3} \end{pmatrix}$ being a generalization of the Pauli matrix to six dimensions (see Ref. [24, Eq. 83]). The reciprocity constraint holds also for complex valued frequencies. From Eq. (B3) we know that $\boldsymbol{\sigma}_z \cdot \left(\partial_i \hat{N}\right)^T \cdot \boldsymbol{\sigma}_z = -\partial_i \hat{N}$. Hence, taking into account that for generic matrices $\mathbf{A}, \mathbf{B}$ the trace has the properties $\mathrm{tr}\{\mathbf{A}\} = \mathrm{tr}\{\mathbf{A}^T\}$ and $\mathrm{tr}\{\mathbf{A}\cdot\mathbf{B}\} = \mathrm{tr}\{\mathbf{B}\cdot\mathbf{A}\}$ it can be shown that $\mathrm{tr}\{\partial_2 \hat{N} \cdot \overline{\mathbf{G}}(\mathbf{r},\mathbf{r}',\omega) \cdot \partial_1 \hat{N} \cdot \partial_\omega \overline{\mathbf{G}}(\mathbf{r}',\mathbf{r},\omega)\} = \mathrm{tr}\{\partial_1 \hat{N} \cdot \overline{\mathbf{G}}(\mathbf{r}',\mathbf{r},\omega) \cdot \partial_2 \hat{N} \cdot \partial_\omega \overline{\mathbf{G}}(\mathbf{r},\mathbf{r}',\omega)\}$. Using this result in Eq. (28) one readily finds that for reciprocal (time-reversal invariant) systems the gap Chern number vanishes, as it should [8, 9, 11].

Equation (29) depends on the Green function values along a contour that is partially contained in the lower-half frequency plane. The derivation of (29) implicitly assumes that it is possible to continue analytically the integrand to the lower half-frequency plane in such a manner that it vanishes for $\omega_{\mathrm{gap}} \pm i\infty$. In Sec. VI, we numerically verify that for electromagnetic continua that is indeed the case. Next, we show that the Chern number may also be written in terms of an integral with the integration path completely contained in the upper-half frequency plane.

To this end, first we note that the term inside the rectangular brackets in Eq. (25) is pure imaginary. Hence, it is simple to check that the integration contour may be restricted to the semi-straight line that joins $\omega_{\mathrm{gap}}$ and $\omega_{\mathrm{gap}} + i\infty$ in the upper-half plane, so that

$$\mathcal{C} = \frac{1}{A_{tot}} \text{Re} \sum_{m,n} \int_{\omega_{gap}}^{\omega_{gap}+i\infty} d\omega \ldots,$$ with the integrand the same as in Eq. (25). Here, "Re" stands for the real-part of a complex number. Hence, following the same steps as before it is found that (compare with Eq. (28))

$$\mathcal{C} = \frac{1}{A_{tot}} \text{Re} \int_{\omega_{gap}}^{\omega_{gap}+i\infty} d\omega \iint dV dV' \left[ \text{tr}\left\{ \partial_2 \hat{N} \cdot \overline{\mathbf{G}}(\mathbf{r},\mathbf{r}',\omega) \cdot \partial_1 \hat{N} \cdot \partial_\omega \overline{\mathbf{G}}(\mathbf{r}',\mathbf{r},\omega) \right\} - 1 \leftrightarrow 2 \right], \quad (30)$$

which gives the Chern number as an integral in the upper-half frequency plane.

### C. Two-dimensional systems

Next, we focus our attention in 2D systems so that the electromagnetic modes are independent of $z$. These systems may regarded as the limit of a 3D waveguide with periodic boundary conditions enforced on the top and bottom walls ($z = 0, d$) and the height of the cavity arbitrarily small, $d \to 0$ (see Fig. 1). In such limit, the $\delta$-distribution in Eq. (8) may be replaced $\delta(\mathbf{r}-\mathbf{r}') \to \frac{1}{d}\delta(x-x')\delta(y-y')$ so that the Green function approaches $\overline{\mathbf{G}}(\mathbf{r},\mathbf{r}',\omega) \to \frac{1}{d}\overline{\mathbf{G}}(\mathbf{r},\mathbf{r}',\omega)$ where $\overline{\mathbf{G}}$ is now the two-dimensional Green function that satisfies $\hat{N} \cdot \overline{\mathbf{G}} = \omega \mathbf{M} \cdot \overline{\mathbf{G}} + i\mathbf{1}\delta(x-x')\delta(y-y')$ with $\partial/\partial z = 0$. Hence, from Eq. (29) the Chern number of 2D-systems is given by:

$$\mathcal{C} = \frac{1}{A_{tot}} \int_{\omega_{gap}-i\infty}^{\omega_{gap}+i\infty} d\omega \iint dS dS' \left[ \text{tr}\left\{ \partial_2 \hat{N} \cdot \overline{\mathbf{G}}(\mathbf{r},\mathbf{r}',\omega) \cdot \partial_1 \hat{N} \cdot \partial_\omega \overline{\mathbf{G}}(\mathbf{r}',\mathbf{r},\omega) \right\} \right], \quad (31)$$

with $dS = dxdy$, $dS' = dx'dy'$ and it is implicit that $\mathbf{r}$ and $\mathbf{r}'$ have now only two components (e.g., $\mathbf{r} = (x,y)$).

Furthermore, for 2D problems often one wishes to restrict the wave polarization to either transverse electric (TE) or transverse magnetic (TM) modes. This can be easily done noting that, for example, TM-waves (with $E_z = 0$ and $H_x = H_y = 0$) stay invariant under the action of the projection operator $\mathbf{1}_{TM} = \text{diag}\{1,1,0,0,0,1\}$, where $\mathbf{1}_{TM}$ is a 6×6 tensor with the indicated diagonal entries. Hence, the gap Chern number associated exclusively with TM waves can be calculated using Eq. (31) with the Green function defined as the solution of

$$\hat{N} \cdot \overline{\mathbf{G}} = \omega \mathbf{M} \cdot \overline{\mathbf{G}} + i\mathbf{1}_{TM} \delta(x-x')\delta(y-y'), \qquad \text{with } \partial/\partial z = 0. \quad (32)$$

## VI. Electromagnetic continuum

Up to now, the analysis of the article is completely general and is valid for a completely generic three-dimensional periodic photonic crystal. To illustrate the application of the developed theory, in the following we specialize it to electromagnetic continua with no intrinsic periodicity [11, 24, 25]. It was shown in Refs. [11, 25] that such material systems may be regarded as topological, provided the nonreciprocal part of the electromagnetic response has a high-frequency spatial cut-off. The introduction of a spatial cut-off frequency requires modifying the material response so that it becomes spatially dispersive [11, 25]. To avoid dealing immediately with such complications, in a first stage we consider only materials with a local response (Sect. VI.A). In a second stage, we further generalize the analysis to materials with a wave vector cut-off (Sect. VI.B). We apply the derived formulas to a magnetized plasma (Sect. VI.C), i.e., to an electric gyrotropic material.

## A. Local material response

The ideas introduced in Sect. V apply with no modifications to electromagnetic continua with no spatial dispersion. For a continuum the Green function is translation invariant and therefore $\overline{\mathbf{G}}(\mathbf{r},\mathbf{r}',\omega) = \overline{\mathbf{G}}(\mathbf{r}-\mathbf{r}',\omega)$. Hence, replacing this result into Eq. (31) and letting $A_{tot} \to \infty$ it is found that:

$$\mathcal{C} = \int_{\omega_{gap}-i\infty}^{\omega_{gap}+i\infty} d\omega \int dS \left[ \text{tr}\left\{ \partial_2 \hat{N} \cdot \overline{\mathbf{G}}(\mathbf{r},\omega) \cdot \partial_1 \hat{N} \cdot \partial_\omega \overline{\mathbf{G}}(-\mathbf{r},\omega) \right\} \right]. \tag{33}$$

In this section, we restrict our attention to TM-polarized waves so that it is implicit that the problem is effectively 2D and that the Green function satisfies Eq. (32). The Green tensor may be formally decomposed as in Eq. (9). For TM-polarized waves, the Green tensor components (3×3 matrices) are of the form: $\mathbf{G}_{EE} = \mathbf{E}^{e1} \otimes \hat{\mathbf{x}} + \mathbf{E}^{e2} \otimes \hat{\mathbf{y}}$, $\mathbf{G}_{ME} = H^{e1}\hat{\mathbf{z}} \otimes \hat{\mathbf{x}} + H^{e2}\hat{\mathbf{z}} \otimes \hat{\mathbf{y}}$, $\mathbf{G}_{EM} = \mathbf{E}^m \otimes \hat{\mathbf{z}}$, and $\mathbf{G}_{MM} = H^m \hat{\mathbf{z}} \otimes \hat{\mathbf{z}}$. Here, $\mathbf{E}^{e,i}$ and $H^{e,i}\hat{\mathbf{z}}$ are respectively the electric and magnetic fields due to an in-plane electric excitation with $\mathbf{j}_e = \delta(x)\delta(y)\hat{\mathbf{u}}_i$ (i=1,2), whereas $\mathbf{E}^m$ and $H^m\hat{\mathbf{z}}$ are the electric and magnetic fields due to an out of plane magnetic excitation with $\mathbf{j}_m = \delta(x)\delta(y)\hat{\mathbf{z}}$.

In practice, for a continuum it is simpler to work in the spectral (Fourier transform in space) domain. Applying the Parseval's theorem to Eq. (33) it is found that:

$$\mathcal{C} = \frac{1}{(2\pi)^2} \int_{\omega_{gap}-i\infty}^{\omega_{gap}+i\infty} d\omega \int d^2\mathbf{k} \left[ \text{tr}\left\{ \partial_2 \hat{N} \cdot \overline{\mathbf{G}}(\mathbf{k},\omega) \cdot \partial_1 \hat{N} \cdot \partial_\omega \overline{\mathbf{G}}(\mathbf{k},\omega) \right\} \right]. \tag{34}$$

where $\overline{\mathbf{G}}(\mathbf{k},\omega)$ is the Fourier transform of the 2D Green function, which may be formally written as:

$$\overline{\mathbf{G}}(\mathbf{k},\omega) = i\left[\hat{N}(\mathbf{k}) - \omega \mathbf{M}(\omega)\right]^{-1} \cdot \mathbf{1}_{TM}, \qquad \text{with } \hat{N}(\mathbf{k}) = \begin{pmatrix} 0 & -\mathbf{k} \times \mathbf{1}_{3\times 3} \\ \mathbf{k} \times \mathbf{1}_{3\times 3} & 0 \end{pmatrix} \qquad (35)$$

and $\mathbf{k} = k_x \hat{\mathbf{x}} + k_y \hat{\mathbf{y}}$. Equations (34)-(35) can be applied in a rather straightforward way to *any* nonreciprocal material without any detailed knowledge of its band structure or eigenfunctions, providing a remarkable simplification as compared to the direct calculation of Chern numbers relying on a gauge dependent Berry potential.

The gap Chern number can also be formally expressed in terms of the fields $\mathbf{E}^{e,i}$, $H^{e,i}\hat{\mathbf{z}}$, $\mathbf{E}^m$, $H^m\hat{\mathbf{z}}$ introduced previously as follows:

$$\mathcal{C} = \frac{1}{(2\pi)^2} \int_{\omega_{gap}-i\infty}^{\omega_{gap}+i\infty} d\omega \int d^2\mathbf{k} \left[ -\tilde{H}^{e2}\partial_\omega \tilde{H}^{e1} - \tilde{H}^m \hat{\mathbf{y}} \cdot \partial_\omega \tilde{\mathbf{E}}^{e1} - \hat{\mathbf{x}} \cdot \tilde{\mathbf{E}}^{e2} \partial_\omega \tilde{H}^m - \hat{\mathbf{x}} \cdot \tilde{\mathbf{E}}^m \hat{\mathbf{y}} \cdot \partial_\omega \tilde{\mathbf{E}}^m \right]. \qquad (36)$$

The tilde hat indicates the fields are Fourier transformed in space.

To consider a specific example, let us suppose that the material response is electric gyrotropic, so that the relative permittivity tensor is of the form:

$$\overline{\varepsilon} = \varepsilon_t \mathbf{1}_t + i\varepsilon_g \hat{\mathbf{z}} \times \mathbf{1}_t + \varepsilon_a \hat{\mathbf{z}} \otimes \hat{\mathbf{z}}, \qquad (37)$$

with $\mathbf{1}_t = \hat{\mathbf{x}} \otimes \hat{\mathbf{x}} + \hat{\mathbf{y}} \otimes \hat{\mathbf{y}}$ and the elements $\varepsilon_{12} = -\varepsilon_{21} = -i\varepsilon_g$ and $\varepsilon_{11} = \varepsilon_{22} = \varepsilon_t$ (the $\varepsilon_{33} = \varepsilon_a$ component is irrelevant for TM-polarized waves). Furthermore, it is supposed that the magnetic response is trivial, $\overline{\mu} = \mathbf{1}_{3\times 3}$, and that $\overline{\xi} = \overline{\zeta} = 0$ (no bianisotropy). The Green function elements ($\mathbf{E}^{e,i}$, $H^{e,i}\hat{\mathbf{z}}$, $\mathbf{E}^m$, $H^m\hat{\mathbf{z}}$) are explicitly evaluated in Appendix C. Substituting (the Fourier transform of) Eq. (C6) into Eq. (36) it is found after integration by parts in frequency of a few terms that:

$$\mathcal{C} = \frac{1}{(2\pi)^2} \int_{\omega_{gap}-i\infty}^{\omega_{gap}+i\infty} d\omega \int d^2\mathbf{k} \frac{2i\varepsilon_g}{\varepsilon_t} \tilde{\Phi} \left[ k^2 \partial_\omega \tilde{\Phi} + \frac{\omega}{c^2} \partial_\omega \left( \omega \varepsilon_{ef} \tilde{\Phi} \right) \right]. \qquad (38)$$

with $\tilde{\Phi} = 1/\left(k^2 - (\omega/c)^2 \varepsilon_{ef}(\omega)\right)$, $k^2 = k_x^2 + k_y^2$, and $\varepsilon_{ef} = \left(\varepsilon_t^2 - \varepsilon_g^2\right)/\varepsilon_t$. Using polar coordinates in the **k**-plane and writing $\omega = \omega_{gap} + i\xi$ one obtains the final result:

$$\mathcal{C} = \frac{-1}{\pi} \int_{-\infty}^{\infty} d\xi \int_0^{\infty} dk\, k\, \frac{\varepsilon_g}{\varepsilon_t} \tilde{\Phi} \left[ k^2 \partial_\omega \tilde{\Phi} + \frac{\omega}{c^2} \partial_\omega \left(\omega \varepsilon_{ef} \tilde{\Phi}\right) \right]_{\omega = \omega_{gap} + i\xi}. \tag{39}$$

It is manifest from the formula that the Chern number can be nonzero only when the material response is nonreciprocal, i.e., when $\varepsilon_g \neq 0$.

### B. Materials with a high-frequency spatial cut-off

The application of topological concepts to electromagnetic continua in general requires the introduction of a high-frequency spatial cut-off $k_{max}$, such that for $k \gg k_{max}$ the material response becomes reciprocal [11, 25]. Conventional material models do not include such an explicit cut-off, but it is physically justified by the fact that real materials have a granular (discrete) nature. One way to mimic a physical spatial cut-off is to modify a given local material response $\mathbf{M}_{loc}(\omega)$ in the following manner [11, 25]:

$$\mathbf{M}(\mathbf{k}, \omega) = \mathbf{M}_\infty + \frac{1}{1 + k^2/k_{max}^2} \left(\mathbf{M}_{loc}(\omega) - \mathbf{M}_\infty\right), \tag{40}$$

so that for $k \ll k_{max}$ the material response is essentially unchanged, whereas for $k \gg k_{max}$ it becomes $\mathbf{M}(\mathbf{k}, \omega) \approx \mathbf{M}_\infty$, i.e., asymptotically the same as that of a reciprocal material. As before, $\mathbf{M}_\infty$ stands for $\mathbf{M}_\infty = \lim_{\omega \to \infty} \mathbf{M}_{loc}(\omega)$, which typically gives the response of the vacuum.

In Appendices D and E, we generalize the theory of the previous subsection to media with a nonlocal response of the form (40). It is shown that as before the Chern number is determined by the photonic Green function in the nonlocal material. Specifically, for a continuum with a cut-off, Eq. (34) becomes $\mathcal{C} = \mathcal{C}_1 + \mathcal{C}_2$ with:

$$\mathcal{C}_1 = \frac{1}{(2\pi)^2} \int_{\omega_{gap}-i\infty}^{\omega_{gap}+i\infty} d\omega \int d^2\mathbf{k} \left(1 + \frac{2k^2}{k^2 + k_{max}^2}\right) \text{tr}\left\{\partial_2 \hat{N} \cdot \overline{\mathbf{G}}(\mathbf{k},\omega) \cdot \partial_1 \hat{N} \cdot \partial_\omega \overline{\mathbf{G}}(\mathbf{k},\omega)\right\}, \quad (41a)$$

$$\mathcal{C}_2 = \frac{-1}{(2\pi)^2} \int d^2\mathbf{k} \frac{1}{k^2 + k_{max}^2} \int_{\omega_{gap}-i\infty}^{\omega_{gap}+i\infty} d\omega \, \omega \left[ k_1 \left(-\text{tr}\left(\mathbf{M}_\infty \cdot \overline{\mathbf{G}} \cdot \partial_2 \hat{N} \cdot \partial_\omega \overline{\mathbf{G}}\right) + \text{tr}\left(\partial_2 \hat{N} \cdot \overline{\mathbf{G}} \cdot \mathbf{M}_\infty \cdot \partial_\omega \overline{\mathbf{G}}\right)\right) \right.$$
$$\left. + k_2 \left(-\text{tr}\left(\partial_1 \hat{N} \cdot \overline{\mathbf{G}} \cdot \mathbf{M}_\infty \cdot \partial_\omega \overline{\mathbf{G}}\right) + \text{tr}\left(\mathbf{M}_\infty \cdot \overline{\mathbf{G}} \cdot \partial_1 \hat{N} \cdot \partial_\omega \overline{\mathbf{G}}\right)\right) \right]$$
(41b)

In the above, $\overline{\mathbf{G}}(\mathbf{k},\omega) = i\left[\hat{N}(\mathbf{k}) - \omega \mathbf{M}(\mathbf{k},\omega)\right]^{-1} \cdot \mathbf{1}_{TM}$ is the spectral 2D Green function for TM-polarized waves (see Appendix F).

For the gyrotropic medium with response (37) the corresponding material with a wave vector cut-off is characterized by a gyrotropic permittivity tensor, $\overline{\varepsilon}(\mathbf{k},\omega) = \varepsilon_t(k,\omega)\mathbf{1}_{3\times 3} + i\varepsilon_g(k,\omega)\hat{\mathbf{z}} \times \mathbf{1}_{3\times 3}$, with components:

$$\varepsilon_t(k,\omega) = 1 + \frac{1}{1 + k^2/k_{max}^2}\left[\varepsilon_{t,\text{loc}}(\omega) - 1\right], \qquad \varepsilon_g(k,\omega) = \frac{\varepsilon_{g,\text{loc}}(\omega)}{1 + k^2/k_{max}^2}, \quad (42)$$

where for simplicity it was assumed that $\varepsilon_{t,\text{loc}}(\infty) = 1$. Hence, the formula for $\overline{\mathbf{G}}(\mathbf{k},\omega)$ is the same as in the local case, except that all the permittivity components become wave vector dependent. Hence, using the results of Appendix C in Eq. (41) it is found after some lengthy but otherwise straightforward calculations that:

$$\mathcal{C}_1 = \frac{-1}{\pi} \int_{-\infty}^{\infty} d\xi \int_0^\infty dk \, k \left(1 + \frac{2k^2}{k^2 + k_{max}^2}\right) \frac{\varepsilon_g}{\varepsilon_t} \tilde{\Phi} \left[k^2 \partial_\omega \tilde{\Phi} + \frac{\omega}{c^2} \partial_\omega \left(\omega \varepsilon_{ef} \tilde{\Phi}\right)\right]_{\omega=\omega_{gap}+i\xi}. \quad (43a)$$

$$C_2 = \frac{-1}{\pi} \int_0^\infty dk \int_{-\infty}^\infty d\xi \, \frac{k^3 \omega}{k^2 + k_{max}^2} \left\{ \frac{1}{\varepsilon_{ef}^2 \omega^2} \left(\frac{\varepsilon_g}{\varepsilon_t}\right)^2 \left( \omega \varepsilon_{ef} \tilde{\Phi} \partial_\omega \left(\frac{\varepsilon_g}{\varepsilon_t}\right) - \frac{\varepsilon_g}{\varepsilon_t} \partial_\omega \left(\omega \varepsilon_{ef} \tilde{\Phi}\right) \right) \right.$$

$$\left. + \frac{1}{c^2} \tilde{\Phi}^2 \left[ 2 \frac{\varepsilon_g}{\varepsilon_t} + \left(\frac{\varepsilon_g}{\varepsilon_t}\right)^2 \partial_\omega \left( \omega \frac{\varepsilon_g}{\varepsilon_t} \right) + \omega \varepsilon_{ef} \partial_\omega \left(\frac{\varepsilon_g}{\varepsilon_t}\right) - \frac{\varepsilon_g}{\varepsilon_t} \partial_\omega (\omega \varepsilon_{ef}) + \frac{c^2}{\varepsilon_{ef}^2 \omega^2} k^2 \partial_\omega \left( \omega \varepsilon_{ef} \frac{\varepsilon_g}{\varepsilon_t} \right) \right] \right\}_{\omega = \omega_{gap} + i\xi}$$

(43b)

where $\tilde{\Phi} = 1/\left[k^2 - (\omega/c)^2 \varepsilon_{ef}(k,\omega)\right]$, $\varepsilon_{ef} = (\varepsilon_t^2 - \varepsilon_g^2)/\varepsilon_t$, and $\varepsilon_t, \varepsilon_g$ must be understood as functions of the wave vector defined as in Eq. (42).

### *C. Magnetized electric plasma*

To illustrate the application of the developed concepts, we consider as an example a magnetized electric plasma. The material has a gyrotropic response as in Eq. (37) with permittivity elements:

$$\varepsilon_t = 1 - \frac{\omega_p^2}{\omega^2 - \omega_0^2}, \qquad \varepsilon_g = \frac{1}{\omega} \frac{\omega_p^2 \omega_0}{\omega_0^2 - \omega^2}. \tag{44}$$

Here, $\omega_0 = -qB_0/m$ is the cyclotron frequency determined by the bias magnetic field $\mathbf{B}_0 = B_0 \hat{\mathbf{z}}$, $q = -e$ is the negative charge of the electrons, $m$ is the effective mass, and $\omega_p$ is the plasma frequency [26].

The band diagrams for propagation in the *xoy* plane are determined from $k^2 = \varepsilon_{ef} \omega^2/c^2$, and hence the dispersion characteristic has rotational symmetry in the *xoy* plane [11, 27, 28]. Figures 3a and 3b show the calculated band structures for the local model [Eq. (44)] and the corresponding nonlocal model [Eq. (42)], respectively, considering a plasma with $\omega_0 = 0.8\omega_p$. Note that both the positive and the negative frequency bands are represented in Fig. 3.

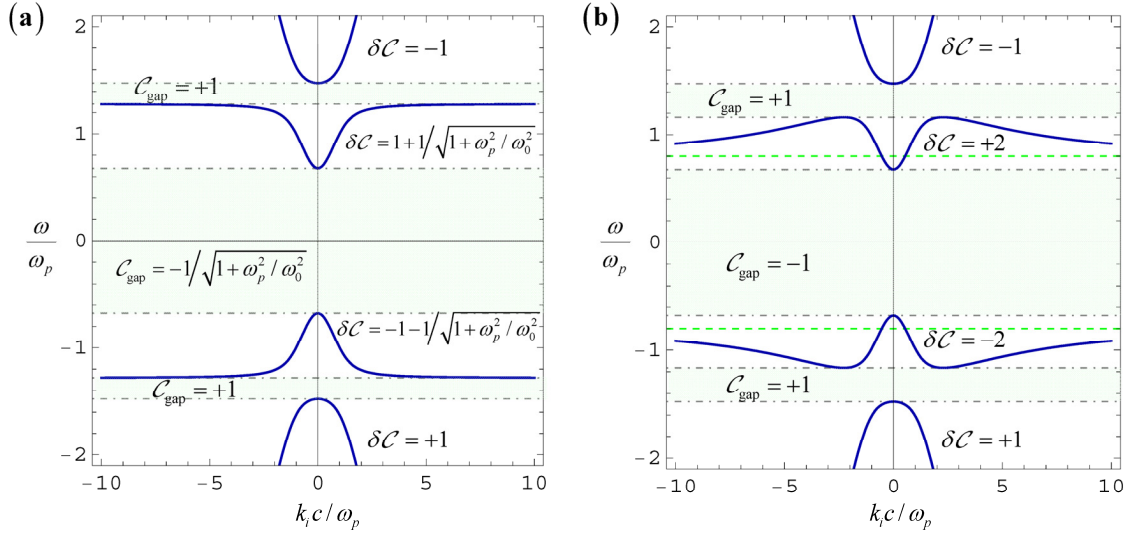

Fig. 3 Band diagram ($\omega$ vs $k_i$, $i=x,y$) of a magnetized electric plasma with $\omega_0 = 0.8\omega_p$ for **a)** local material response. **b)** material response with the spatial cut-off $k_{max} = 5\omega_p/c$. The shaded areas indicate the electromagnetic band gaps. The Chern numbers ($\delta\mathcal{C}$) of the individual bands are given in the insets and were calculated using the theory of Ref. [11]. The gap Chern numbers ($\mathcal{C}_{gap}$) determined by the photonic Green function using Eqs. (39) (in the local case) and (43) (in the nonlocal case) are also indicated in the figure, and agree with the sum of the individual Chern numbers of the bands below the gap.

As seen, there are 3 distinct complete band gaps (2 complete band gaps if one considers only positive frequency branches). The two positive frequency branches may be regarded as the result of the perturbation (by the bias magnetic field) of the transverse and longitudinal wave branches supported by a standard non-magnetized plasma with a Drude-type response, $\varepsilon = 1 - \omega_p^2/\omega^2$. The bias magnetic field opens a band gap in between these two branches. The local and nonlocal dispersions are nearly coincident for $k \ll k_{max} = 5\omega_p/c$. However, in the nonlocal model the waves associated with the low positive-frequency branch become backward for $k > k_{max}$ and asymptotically, as $k \to \infty$, this branch approaches $|\omega_0|/c$ (dashed green line in Fig. 3b).

The individual Chern numbers associated with each band ($\delta\mathcal{C}$) can be found directly from the electromagnetic eigenstates using the formalism of Ref. [11]. They are indicated as insets in Fig. 3. As seen, consistent with Ref. [11], without a spatial cut-off (Fig. 3a) the Chern numbers may be non-integer. For example, the Chern number of the low positive-frequency branch is $\delta\mathcal{C}=1+1/\sqrt{1+\omega_p^2/\omega_0^2}$ [27]. Indeed, the topological classification of a continuum is strictly valid only in presence of the spatial cut-off. The Chern numbers of the negative frequency bands differ always by a minus sign from the Chern numbers of the corresponding positive frequency bands. Furthermore, the Chern numbers flip sign when the direction of the bias magnetic field is reversed.

It is interesting to note that the sum of all positive-frequency bands Chern numbers is nonzero (+1) when the spatial cut-off is enforced. This property remains true for any $\omega_0 > 0$, i.e., any positive bias magnetic field. For $|\omega_0| \ll \omega_p$, the low-frequency band-gap is determined by $-|\omega_0| < \omega < |\omega_0|$. Hence, in the limit $\omega_0 \to 0^+$ this band gap closes and there is an interaction between the positive and negative frequency branches resulting in an exchange of topological charge, such that for $\omega_0 = 0$ (reciprocal response) the Chern numbers of the positive-frequency branches are precisely zero, as it should be. A similar topological interaction between the positive and negative frequency branches has been discussed in Refs. [29, 30], but typically the role of the negative frequency modes is overlooked in the literature.

Using Eqs. (39) and (43) for the local and nonlocal cases, respectively, it is straightforward to evaluate the gap Chern numbers for each of the band gaps. The result of the calculations is indicated in the insets of Fig. 3, and agrees perfectly with what is found by explicitly summing the individual Chern number contributions ($\mathcal{C}_{gap} = \sum\limits_{\substack{\text{low-freq} \\ \text{bands}}} \delta \mathcal{C}_i$). Note that this property holds even in the local case (without spatial cut-off) when the Chern number is not necessarily an integer. This result validates the theory developed here.

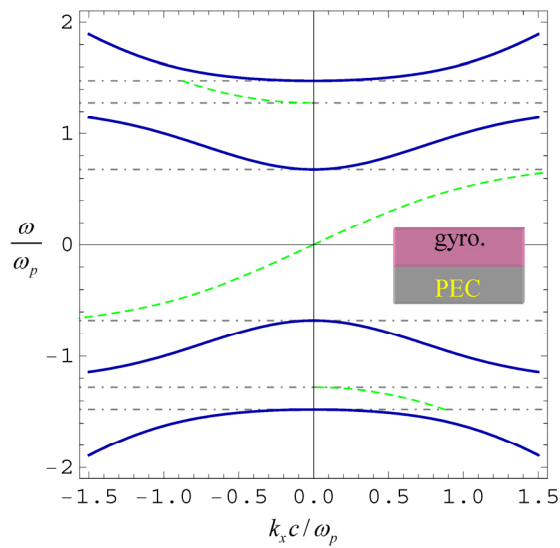

Fig. 4 Band diagram of the edge modes (dashed green lines) supported by an interface ($y$=0) between the gyrotropic material with $\omega_0 = 0.8\omega_p$ (region $y$>0) and PEC material (region $y$<0). The edge modes propagate along the $x$-direction. The solid blue lines represent the band diagram of the bulk gyrotropic material. The dispersion diagrams are obtained without an explicit spatial cut-off ($k_{max} \to \infty$).

We would like to highlight that it is absolutely essential to consider the contribution of the negative frequency branches in the calculation of the gap Chern number. Indeed, a summation of Chern numbers restricted to positive frequency bands would yield a gap Chern number equal to 0 for the first (low-frequency) band gap, and +2, for the second (high-frequency) gap. From the bulk-edge correspondence this would imply that an interface between the magnetized plasma and a trivial insulator should support 0 and 2 unidirectional edge modes, in each of the band gaps. However, as illustrated in Fig. 4 for the case of an interface with a perfect electric conductor, the number of edge modes in each gap is precisely 1. This result agrees with the gap Chern numbers reported in Fig. 3b, and confirms the importance of taking into account the contributions of the negative frequency bands. The edge modes are calculated as explained in Refs. [11, 25]. For the limitations on the application of the bulk-edge correspondence to electromagnetic continua the reader is referred to Ref. [25].

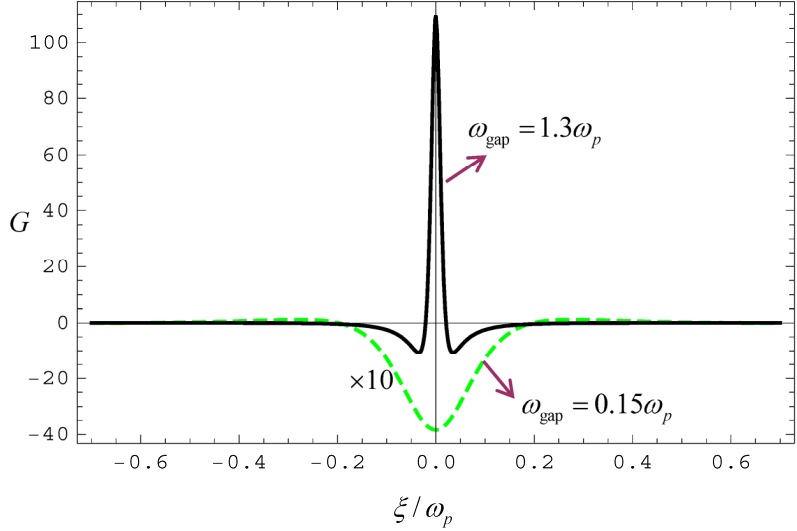

Fig. 5 Normalized integrand $G\left(\omega_{gap} + i\xi\right)$ of the Chern number integral [Eq. (39)] as a function of the imaginary part of the angular frequency ($\xi$) at (*i*) the low positive-frequency band-gap ($\omega_{gap} = 0.15\omega_p$) and at (*ii*) the high positive-frequency band-gap ($\omega_{gap} = 1.3\omega_p$). The cyclotron frequency is $\omega_0 = 0.8\omega_p$.

It is useful to give an idea of the asymptotic behavior of the integrand of Eqs. (39) and (43) and thus of the Green function behavior in the complex frequency plane. To this end, we depict in Fig. 5 the dimensionless function $G\left(\omega_{gap} + i\xi\right)$ as a function of $\xi$, with $G\left(\omega_{gap} + i\xi\right)$ defined such that Eq. (39) can be rewritten as

$$\mathcal{C} = \int_{-\infty}^{\infty} d\xi \frac{1}{\omega_p} \mathrm{Re}\{G(\omega_{\mathrm{gap}} + i\xi)\}.$$ As seen, the integrand is peaked near $\xi = 0$ (real-valued frequency $\omega = \omega_{\mathrm{gap}}$), and decays exponentially fast as $\xi \to \pm\infty$, exhibiting an even symmetry in this example. The integrand has no singularities because the integration path lies in the band-gap and hence does not cross any poles. Clearly, the Chern number is determined by the behavior of the photonic Green function near the real frequency axis.

## VII. Summary

We established a link between the gap Chern number of a generic bianisotropic inhomogeneous nonreciprocal periodic optical platform and the photonic Green function. The developed theory applies to fully three-dimensional closed systems, such that the energy can only flow along directions parallel to the *xoy* plane. Furthermore, the formalism can also be applied to electromagnetic continua with no intrinsic periodicity and to electromagnetic continua with a spatial cut-off.

The main result of the article is given by Eq. (29) and establishes that the gap Chern number can be written as an integral in the complex frequency plane of the photonic Green function and of its derivative in frequency. The integration is along a path parallel to the imaginary frequency axis. The Green function has no singularities in the integration path and its asymptotic behavior ensures that the integral that gives the Chern number decays exponentially fast. Furthermore, our theory does not require any explicit knowledge of the detailed photonic band structure or of the Bloch eigenstates. Thus, Eq. (29) may be useful to compute the Chern numbers of complex photonic platforms using numerical methods. Note that the Green function in the complex-frequency plane can be numerically evaluated using standard finite-difference-frequency-domain approaches.

Moreover, our theory highlights the key role of negative frequency modes, which are often disregarded in the calculation of gap Chern numbers.

Most of our derivation can be readily extended to electronic systems (as mentioned in the introduction it is not equivalent to that of Ref. [14]), but it is worth pointing out a little difference: unlike in optics, in the electronic case the operators $\partial_i \hat{H}_g$ typically depend explicitly on the wave vector. Due to this reason, Eq. (29) (written in terms of the spatial domain Green function) does not hold in the electronic case. The electronic gap Chern number can however be related to Bloch-periodic Green functions, but a detailed discussion is out of the intended scope of this study.

The deep link between the gap Chern number and the photonic Green function raises intriguing questions about the origin of the topological properties of photonic systems. Furthermore, somewhat similar to the zero-point energy of a system [31, 32], the Chern number is written in terms of an integral over imaginary frequencies, and this suggests that it may be related to the quantum expectation of some physical quantity. These fundamental connections will be discussed elsewhere.

**Acknowledgements:** This work is supported in part by Fundação para a Ciência e a Tecnologia grant number PTDC/EEI-TEL/4543/2014 and by Instituto de Telecomunicações under project UID/EEA/50008/2013.

## Appendix A: Modal expansions in dispersive media

The spectral theorem guarantees that an arbitrary state vector $\mathbf{Q}(\mathbf{r})$ can be expanded in normal modes of oscillation of the Hermitian operator $\hat{H}_g = \mathbf{M}_g^{-1} \cdot \hat{L}$. Hence, if $\mathbf{Q}_1, \mathbf{Q}_2, \ldots$ determine an orthogonal basis of eigenmodes of $\hat{H}_g$ associated with the eigenfrequencies $\omega_1, \omega_2, \ldots$, i.e., $\hat{H}_g \mathbf{Q}_n = \omega_n \mathbf{Q}_n$, it is possible to write:

$$\mathbf{Q}(\mathbf{r}) = \sum_n \mathbf{Q}_n(\mathbf{r}) c_n \qquad \text{with} \qquad c_n = \langle \mathbf{Q}_n | \mathbf{Q} \rangle. \tag{A1}$$

It is implicit that the eigenmodes are normalized such that $\langle \mathbf{Q}_n | \mathbf{Q}_m \rangle = \delta_{n,m}$. The completeness of the basis set implies that

$$\delta(\mathbf{r} - \mathbf{r}') \mathbf{1}_g = \frac{1}{2} \sum_n \mathbf{Q}_n(\mathbf{r}) \otimes \mathbf{Q}_n^*(\mathbf{r}') \cdot \mathbf{M}_g(\mathbf{r}'), \tag{A2}$$

where $\otimes$ represents the tensor product of two vectors, and $\mathbf{1}_g$ is the identity tensor with the same dimension as $\mathbf{M}_g$. Let $\mathbf{f}_n$ be the electromagnetic field component of $\mathbf{Q}_n$, so that $\mathbf{Q}_n = \begin{pmatrix} \mathbf{f}_n & \mathbf{Q}_n^{(1)} & \ldots \end{pmatrix}^T$. Then, the result (11) and the normalization condition $\langle \mathbf{Q}_n | \mathbf{Q}_m \rangle = \delta_{n,m}$ show that the modes $\mathbf{f}_n$ are normalized as in Eq. (13) of the main text. Furthermore, Eq. (A2) implies the electromagnetic modes satisfy the completeness relation (14). In particular, by multiplying both sides of Eq. (14) by a generic electromagnetic field distribution $\mathbf{f}(\mathbf{r})$ and integrating the resulting equation over $\mathbf{r}'$, we obtain the modal expansion:

$$\mathbf{f}(\mathbf{r}) = \sum_n \mathbf{f}_n(\mathbf{r}) c_n, \qquad \text{with} \qquad c_n = \frac{1}{2} \int_V d^3\mathbf{r}' \mathbf{f}_n^*(\mathbf{r}') \cdot \mathbf{M}_\infty(\mathbf{r}') \cdot \mathbf{f}(\mathbf{r}'). \tag{A3}$$

Crucially, in a dispersive system the expansion coefficients $c_n$ are *not* unique, i.e., the same field distribution $\mathbf{f}(\mathbf{r})$ can be obtained with different sets of coefficients $c_n$. To illustrate this property, we take $\mathbf{f}(\mathbf{r}) = \mathbf{f}_m(\mathbf{r})$ which evidently satisfies $\mathbf{f}(\mathbf{r}) = \sum_n \mathbf{f}_n(\mathbf{r}) c_n$ with $c_n = \delta_{n,m}$. However, for a dispersive system the coefficients obtained from Eq. (A3) with $\mathbf{f}(\mathbf{r}) = \mathbf{f}_m(\mathbf{r})$ typically differ from $c_n = \delta_{n,m}$ because the generalized orthogonality conditions are as in Eq. (13). The lack uniqueness of the expansion coefficients may be understood by noting that the $\{\mathbf{f}_n\}_{n=1,2,...}$ set is a projection of the original $\{\mathbf{Q}_n\}_{n=1,2,...}$ basis of the augmented space, and thereby it "over" spans the electromagnetic space. In contrast, modal expansions [Eq. (A1)] in the augmented space are unique.

Let $\mathbf{f}(\mathbf{r})$ be some solution of the Maxwell's equations (10) with oscillation frequency $\omega$, and let $\mathbf{Q}$ be the corresponding solution of the generalized time-harmonic problem, $\hat{L} \cdot \mathbf{Q} = \omega \mathbf{M}_g \cdot \mathbf{Q} + i \mathbf{j}_g$. Then, using Eqs. (A1) and (11) it follows that the expansion coefficients may be written as:

$$c_n = \frac{1}{2} \int_V d^3\mathbf{r}\, \mathbf{f}_n^*(\mathbf{r}) \cdot \left[ \frac{\omega_n \mathbf{M}(\mathbf{r}, \omega_n) - \omega \mathbf{M}(\mathbf{r}, \omega)}{\omega_n - \omega} \right] \cdot \mathbf{f}(\mathbf{r})$$
$$= \frac{1}{2} \int_V d^3\mathbf{r}\, \frac{\left[\hat{N}\mathbf{f}_n\right]^* \cdot \mathbf{f} - \mathbf{f}_n^* \cdot \left(\hat{N}\mathbf{f} - i\mathbf{j}\right)}{\omega_n - \omega}. \quad (A4)$$

In the second identity, we used Eq. (12) and $\mathbf{M} = \mathbf{M}^\dagger$. Noting that the differential operator $\hat{N}$ is Hermitian with respect to the canonical inner product, we readily conclude that $c_n$ may be written as in Eq. (15), and thereby the electromagnetic field has the expansion given in the same equation.

## Appendix B: The Chern number modal expansion

In the following, we derive Eq. (21) of the main text. To this end, first we note that because the eigenstates of the augmented problem form a complete set, the Berry curvature [Eq. (19)] can be rewritten as

$$\mathcal{F}_{n\mathbf{k}} = \sum_{m\mathbf{k} \neq n\mathbf{k}} i \left[ \langle \partial_1 \mathbf{Q}_{n\mathbf{k}} | \mathbf{Q}_{m\mathbf{k}} \rangle \langle \mathbf{Q}_{m\mathbf{k}} | \partial_2 \mathbf{Q}_{n\mathbf{k}} \rangle - \langle \partial_2 \mathbf{Q}_{n\mathbf{k}} | \mathbf{Q}_{m\mathbf{k}} \rangle \langle \mathbf{Q}_{m\mathbf{k}} | \partial_1 \mathbf{Q}_{n\mathbf{k}} \rangle \right]. \quad (B1)$$

Furthermore, from Eq. (17) it is seen that

$$\partial_\mathbf{k} \hat{H}_g \cdot \mathbf{Q}_{n\mathbf{k}} + \hat{H}_g \cdot \partial_\mathbf{k} \mathbf{Q}_{n\mathbf{k}} = \omega_{n\mathbf{k}} \partial_\mathbf{k} \mathbf{Q}_{n\mathbf{k}} + \partial_\mathbf{k} \omega_{n\mathbf{k}} \mathbf{Q}_{n\mathbf{k}},$$

so that for $m\mathbf{k} \neq n\mathbf{k}$ one has $\langle \mathbf{Q}_{m\mathbf{k}} | \partial_\mathbf{k} \hat{H}_g | \mathbf{Q}_{n\mathbf{k}} \rangle = (\omega_{n\mathbf{k}} - \omega_{m\mathbf{k}}) \langle \mathbf{Q}_{m\mathbf{k}} | \partial_\mathbf{k} \mathbf{Q}_{n\mathbf{k}} \rangle$. Thus, it is possible to write [23]:

$$\mathcal{F}_{n\mathbf{k}} = \sum_{m\mathbf{k} \neq n\mathbf{k}} \frac{i}{(\omega_{n\mathbf{k}} - \omega_{m\mathbf{k}})^2} \left[ \langle \mathbf{Q}_{n\mathbf{k}} | \partial_1 \hat{H}_g | \mathbf{Q}_{m\mathbf{k}} \rangle \langle \mathbf{Q}_{m\mathbf{k}} | \partial_2 \hat{H}_g | \mathbf{Q}_{n\mathbf{k}} \rangle - 1 \leftrightarrow 2 \right]. \tag{B2}$$

The term "$1 \leftrightarrow 2$" is defined as in the main text. The operator $\partial_i \hat{H}_g$ stands for $\frac{\partial}{\partial k_i} \left[ \hat{H}_g(\mathbf{r}, -i\nabla + \mathbf{k}) \right]$. Taking into account that $\hat{H}_g = \mathbf{M}_g^{-1} \cdot \hat{L}$ and the definitions of Eq. (5) it is simple to check that $\partial_i \hat{H}_g = \mathbf{M}_g^{-1} \cdot \partial_i \hat{L}$ with

$$\partial_i \hat{L} = \begin{pmatrix} \partial_i \hat{N} & \mathbf{0} & \dots \\ \mathbf{0} & \mathbf{0} & \dots \\ \dots & \dots & \dots \end{pmatrix}, \qquad \text{where } \partial_i \hat{N} = \begin{pmatrix} \mathbf{0} & -\hat{\mathbf{u}}_i \times \mathbf{1}_{3\times 3} \\ \hat{\mathbf{u}}_i \times \mathbf{1}_{3\times 3} & \mathbf{0} \end{pmatrix} \tag{B3}$$

and $\hat{\mathbf{u}}_i$ is a unit vector along the $i$-th direction. Hence, the term $\langle \mathbf{Q}_{m\mathbf{k}} | \partial_i \hat{H}_g | \mathbf{Q}_{n\mathbf{k}} \rangle$ only depends on the electromagnetic field components of the state vectors.

In order to write the Chern number (20) in terms of a discrete summation, it is supposed without loss of generality that the unit cell has dimensions $a_x \times a_y$, so that the Brioullin zone corresponds to the rectangular region $\left[-\frac{\pi}{a_x}, \frac{\pi}{a_x}\right] \times \left[-\frac{\pi}{a_y}, \frac{\pi}{a_y}\right]$. Furthermore, we consider that $V$ (the "cavity") is a region with $N_x \times N_y$ cells terminated with periodic boundaries. Then, we may write $d^2\mathbf{k} \approx \frac{(2\pi)^2}{A_{tot}}$ with $A_{tot} = (a_x N_x) \times (a_y N_y)$ the transverse area of the volume $V$, so that Eq. (20) becomes:

$$\mathcal{C} = \lim_{A_{tot} \to \infty} \frac{2\pi}{A_{tot}} \sum_{n \in F} \mathcal{F}_n, \tag{B4}$$

where $\mathcal{F}_n$ is now given by (compare with Eq. (B2)):

$$\mathcal{F}_n = \sum_{m \neq n} i \frac{1}{(\omega_n - \omega_m)^2} \left[ \langle \mathbf{Q}_n | \partial_1 \hat{H}_g | \mathbf{Q}_m \rangle \langle \mathbf{Q}_m | \partial_2 \hat{H}_g | \mathbf{Q}_n \rangle - 1 \leftrightarrow 2 \right]. \tag{B5}$$

The summation in Eq. (B4) is over the modes of the "cavity" (the volume $V$) with $\omega_n < \omega_{gap}$, whereas the summation in (B5) is over all the modes. As explained in the main text, $\{\mathbf{Q}_n\}_{n=1,2,...}$ may be taken as the "full" modes of the cavity rather than the spatial envelopes.

The Chern number (B4) can be decomposed into two parcels, $\mathcal{C} = \frac{2\pi}{A_{tot}} \sum_{\substack{m \neq n, \\ n \in F \\ m \in F}} \theta_{m,n} + \frac{2\pi}{A_{tot}} \sum_{\substack{n \in F \\ m \in E}} \theta_{m,n}$, where the definition of $\theta_{m,n}$ should be evident by inspection of Eq. (B5). Clearly, the first term vanishes due to the anti-symmetry of $\theta_{m,n}$ with respect to interchanging the indices $m$ and $n$. This proves that the Chern number is due to the interaction of filled ($F$) and empty ($E$) bands, and thus Eq. (21) of the main text follows.

To conclude we note that the anti-symmetry of $\theta_{m,n}$ implies that the total Chern number (i.e., the sum of the Chern numbers of all the individual bands) must vanish: $0 = \frac{2\pi}{A_{tot}} \sum_{m,n} \theta_{m,n}$. This identity shows that $0 = \frac{2\pi}{A_{tot}} \sum_{\substack{n \in F \\ m \in E}} \theta_{m,n} + \frac{2\pi}{A_{tot}} \sum_{\substack{n \in E \\ m \in F}} \theta_{m,n}$ because as previously noted $0 = \sum_{\substack{m \neq n, \\ n \in F \\ m \in F}} \theta_{m,n} = \sum_{\substack{m \neq n, \\ n \in E \\ m \in E}} \theta_{m,n}$. Hence, the gap Chern number in Eq. (21) may also be expressed as $\mathcal{C} = -\frac{2\pi}{A_{tot}} \sum_{\substack{n \in E, \\ m \in F}} (...)$, with the generic term of summation the same as in Eq. (21).

## Appendix C: The 2D Green function for a gyrotropic material

In this Appendix, we obtain explicit expressions for the 2D Green function components ($\mathbf{E}^{e,i}$, $H^{e,i}\hat{\mathbf{z}}$, $\mathbf{E}^m$ and $H^m\hat{\mathbf{z}}$) introduced in the main text for the case of an electric gyrotropic material with permittivity tensor as in Eq. (37). These fields can be found from the solution of the Maxwell's equations:

$$\nabla \times \mathbf{H} = -i\omega\varepsilon_0 \overline{\varepsilon} \cdot \mathbf{E} + \mathbf{j}_{e0}\delta(\mathbf{r}_\parallel), \qquad \nabla \times \mathbf{E} = i\omega\mu_0\mathbf{H} - \hat{\mathbf{z}} j_{m0}\delta(\mathbf{r}_\parallel), \qquad (C1)$$

with $\partial/\partial z = 0$ and $\delta(\mathbf{r}_\parallel) = \delta(x)\delta(y)$. Specifically, $\mathbf{E}^{e,i}$, $H^{e,i}\hat{\mathbf{z}}$ are the electromagnetic fields for $\mathbf{j}_{e0} = \hat{\mathbf{u}}_i$ ($i=1,2$) and $j_{m0} = 0$, whereas $\mathbf{E}^m$ and $H^m\hat{\mathbf{z}}$ are the fields for $\mathbf{j}_{e0} = 0$ and $j_{m0} = 1$.

For TM-polarization $\mathbf{H} = H_z\hat{\mathbf{z}}$, and thus the electric field excited by the sources is:

$$-i\omega\varepsilon_0\mathbf{E} = \overline{\varepsilon}^{-1} \cdot (\nabla H_z \times \hat{\mathbf{z}}) - \overline{\varepsilon}^{-1} \cdot \mathbf{j}_{e0}\delta(\mathbf{r}_\parallel), \qquad (C2)$$

where $\overline{\varepsilon}^{-1} = \frac{1}{\varepsilon_{ef}}\left(\mathbf{1}_t - i\frac{\varepsilon_g}{\varepsilon_t}\hat{\mathbf{z}} \times \mathbf{1}_t\right) + \frac{1}{\varepsilon_a}\hat{\mathbf{z}} \otimes \hat{\mathbf{z}}$ is the inverse permittivity tensor and $\varepsilon_{ef}$ is defined as in the main text. Substituting the above formula into Faraday's equation it is found after some manipulations that $H_z$ must satisfy:

$$\nabla^2 H_z + \left(\frac{\omega}{c}\right)^2 \varepsilon_{ef} H_z = \nabla\delta(\mathbf{r}_\parallel) \cdot \left[\hat{\mathbf{z}} \times \varepsilon_{ef} \overline{\varepsilon}^{-1} \cdot \mathbf{j}_{e0}\right] - i\omega\varepsilon_0\varepsilon_{ef} j_{m0}\delta(\mathbf{r}_\parallel) \qquad (C3)$$

The solution of this equation is:

$$H_z = -\nabla\Phi \cdot \left[\hat{\mathbf{z}} \times \varepsilon_{ef}\overline{\varepsilon}^{-1} \cdot \mathbf{j}_{e0}\right] + i\omega\varepsilon_0\varepsilon_{ef} j_{m0}\Phi \qquad (C4)$$

where $\Phi$ satisfies $\nabla^2 \Phi + \left(\dfrac{\omega}{c}\right)^2 \varepsilon_{ef} \Phi = -\delta(\mathbf{r}_\parallel)$, and thereby may be explicitly written in terms of Hankel functions. From Eqs. (C2) and (C4) it follows that the Green function components are:

$$\dfrac{H^m}{i\omega\varepsilon_0} = \varepsilon_{ef}\Phi, \qquad \mathbf{E}^m = -\varepsilon_{ef}\overline{\overline{\varepsilon}}^{-1} \cdot (\nabla\Phi \times \hat{\mathbf{z}}) \qquad (C5a)$$

$$H^{e,i} = -\nabla\Phi \cdot \left[\hat{\mathbf{z}} \times \varepsilon_{ef}\overline{\overline{\varepsilon}}^{-1} \cdot \hat{\mathbf{u}}_i\right], \quad i\omega\varepsilon_0\mathbf{E}^{e,i} = -\overline{\overline{\varepsilon}}^{-1} \cdot \left(\nabla\times\nabla\times\left(\varepsilon_{ef}\overline{\overline{\varepsilon}}^{-1} \cdot \hat{\mathbf{u}}_i\Phi\right)\right) + \overline{\overline{\varepsilon}}^{-1} \cdot \hat{\mathbf{u}}_i\delta(\mathbf{r}_\parallel)$$

(C5b)

In particular, the terms that determine the Chern number in Eq. (36) are:

$$\dfrac{H^m}{i\omega\varepsilon_0} = \varepsilon_{ef}\Phi, \qquad \begin{cases}\hat{\mathbf{x}} \cdot \mathbf{E}^m \\ \hat{\mathbf{y}} \cdot \mathbf{E}^m\end{cases} = \begin{cases}-\partial_y\Phi + \dfrac{i\varepsilon_g}{\varepsilon_t}\partial_x\Phi \\ +\partial_x\Phi + \dfrac{i\varepsilon_g}{\varepsilon_t}\partial_y\Phi\end{cases} \qquad (C6a)$$

$$\begin{cases}H^{e,1} \\ H^{e,2}\end{cases} = \begin{cases}-\partial_y\Phi - \dfrac{i\varepsilon_g}{\varepsilon_t}\partial_x\Phi \\ +\partial_x\Phi - \dfrac{i\varepsilon_g}{\varepsilon_t}\partial_y\Phi\end{cases}, \qquad i\omega\varepsilon_0\begin{cases}\hat{\mathbf{y}} \cdot \mathbf{E}^{e,1} \\ \hat{\mathbf{x}} \cdot \mathbf{E}^{e,2}\end{cases} = -\dfrac{1}{\varepsilon_t}\partial_x\partial_y\Phi \pm \dfrac{i\varepsilon_g}{\varepsilon_t}\left(\dfrac{\omega}{c}\right)^2\Phi \qquad (C6b)$$

The fields Fourier transforms are found with the usual rules $\partial_x \to ik_x$, $\partial_y \to ik_y$ and $\Phi \to \tilde{\Phi} = 1/\left(k^2 - (\omega/c)^2\varepsilon_{ef}\right)$.

## Appendix D: Electromagnetic continuum with a cut-off

Here, we generalize the theory of Sect. VI.A to an electromagnetic continuum with a spatial cut-off. The system is assumed uniform along the $z$-direction. For a material with response as in Eq. (40), it was shown in Ref. [11] that the electromagnetic modes (plane waves with propagation factor $\mathbf{f}_{n\mathbf{k}}e^{i\mathbf{k}\cdot\mathbf{r}}$) can be found from the solution of an augmented problem of the form $\hat{H}_{nl}(\mathbf{k})\cdot \mathbf{Q}_{n\mathbf{k}} = \omega_{n\mathbf{k}} \mathbf{Q}_{n\mathbf{k}}$ with state vector $\mathbf{Q}_{n\mathbf{k}} = \begin{pmatrix} \mathbf{f}_{n\mathbf{k}} & \mathbf{Q}_{n\mathbf{k}}^{(1)} & ... \end{pmatrix}^T$.

Here, $\hat{H}_{nl} = \mathbf{M}_g^{-1} \cdot \hat{L}_{nl}(\mathbf{k})$ with $\hat{L}_{nl}$ given by

$$\hat{L}_{nl} = \begin{pmatrix} \hat{N}(\mathbf{k}) + \dfrac{1}{1+k^2/k_{max}^2} \sum_{\alpha} \text{sgn}(\omega_{p,\alpha}) \mathbf{A}_{\alpha}^2 & \dfrac{1}{(1+k^2/k_{max}^2)^{1/2}} |\omega_{p,1}|^{1/2} \mathbf{A}_1 & \dfrac{1}{(1+k^2/k_{max}^2)^{1/2}} |\omega_{p,2}|^{1/2} \mathbf{A}_2 & ... \\ \dfrac{1}{(1+k^2/k_{max}^2)^{1/2}} |\omega_{p,1}|^{1/2} \mathbf{A}_1 & \omega_{p,1}\mathbf{1} & 0 & ... \\ \dfrac{1}{(1+k^2/k_{max}^2)^{1/2}} |\omega_{p,2}|^{1/2} \mathbf{A}_2 & 0 & \omega_{p,2}\mathbf{1} & ... \\ ... & ... & ... & ... \end{pmatrix}$$

(D1)

where $\omega_{p,\alpha}$ and $\mathbf{A}_{\alpha}$ are the same coefficients as in the local case. Furthermore, $\mathbf{M}_g$ is defined as in Eq. (5) and $\hat{N}(\mathbf{k})$ as in Eq. (35). For a fixed wave vector, the operator $\hat{H}_{nl}$ is Hermitian with respect to the weighted inner product (6). Hence, similar to Sect. IV, it is possible to introduce a Berry potential and a gap Chern number. The spatial-cut off ensures that the gap Chern number is an integer [11].

Following the same sequence of steps as in Sects. IV and V, equation (25) may be generalized to the case of a continuum with a spatial cut-off as follows:

$$\mathcal{C} = \dfrac{1}{2A_{tot}} \sum_{m\mathbf{k},n\mathbf{k}} \int_{\omega_{gap}-i\infty}^{\omega_{gap}+i\infty} d\omega \dfrac{1}{(\omega-\omega_{m\mathbf{k}})^2} \dfrac{1}{\omega-\omega_{n\mathbf{k}}} \Big[ \langle \tilde{\mathbf{Q}}_{n\mathbf{k}} | \partial_1 \hat{H}_{nl} | \tilde{\mathbf{Q}}_{m\mathbf{k}} \rangle \langle \tilde{\mathbf{Q}}_{m\mathbf{k}} | \partial_2 \hat{H}_{nl} | \tilde{\mathbf{Q}}_{n\mathbf{k}} \rangle - 1 \leftrightarrow 2 \Big].$$

(D2)

In the above, $\partial_i \hat{H}_{nl} = \partial \hat{H}_{nl}/\partial k_i$ is the derivative of the operator $\hat{H}_{nl}$ with respect to the wave vector. The modes are normalized such that $\langle \tilde{\mathbf{Q}}_{n\mathbf{k}} | \tilde{\mathbf{Q}}_{n\mathbf{k}} \rangle = 1$. Taking the limit $A_{tot} \to \infty$, and using the definition of the weighted inner product (6) we find:

$$\mathcal{C} = \frac{1}{8} \frac{1}{(2\pi)^2} \int d^2\mathbf{k} \sum_{m,n} \int_{\omega_{gap}-i\infty}^{\omega_{gap}+i\infty} d\omega \frac{1}{(\omega-\omega_{m\mathbf{k}})^2} \frac{1}{\omega-\omega_{n\mathbf{k}}} \left[ \mathbf{Q}_{n\mathbf{k}}^* \cdot \partial_1 \hat{L}_{nl} \cdot \mathbf{Q}_{m\mathbf{k}} \mathbf{Q}_{m\mathbf{k}}^* \cdot \partial_2 \hat{L}_{nl} \cdot \mathbf{Q}_{n\mathbf{k}} - 1 \leftrightarrow 2 \right]$$

(D3)

with $\mathbf{Q}_{n\mathbf{k}} = \tilde{\mathbf{Q}}_{n\mathbf{k}} \sqrt{V_{tot}}$ and $V_{tot}$ the volume of the considered cavity. The modes $\mathbf{Q}_{n\mathbf{k}}$ satisfy the normalization condition $\frac{1}{2} \mathbf{Q}_{n\mathbf{k}}^* \cdot \mathbf{M}_g \cdot \mathbf{Q}_{n\mathbf{k}} = 1$. It is proven in Appendix E that:

$$\mathbf{Q}_{n\mathbf{k}}^* \cdot \partial_i \hat{L}_{nl} \cdot \mathbf{Q}_{m\mathbf{k}} = \mathbf{f}_{n\mathbf{k}}^* \cdot \partial_i \hat{N} \cdot \mathbf{f}_{m\mathbf{k}} + \frac{2k_i}{k^2 + k_{max}^2} \left[ \mathbf{f}_{n\mathbf{k}}^* \cdot \left( \hat{N}(\mathbf{k}) - \mathbf{M}_\infty \frac{\omega_{m\mathbf{k}} + \omega_{n\mathbf{k}}}{2} \right) \cdot \mathbf{f}_{m\mathbf{k}} \right]. \quad \text{(D4)}$$

Thus, the Chern number can be written simply as a function of the electromagnetic components ($\mathbf{f}_{n\mathbf{k}}$) of the modes ($\mathbf{Q}_{n\mathbf{k}}$) of the augmented problem.

The next step is to write the Chern number using the spectral Green function $\overline{\mathbf{G}}(\mathbf{k}, \omega)$ of the nonlocal problem, defined as:

$$\overline{\mathbf{G}}(\mathbf{k}, \omega) = \frac{-i}{2} \sum_n \frac{1}{\omega - \omega_{n\mathbf{k}}} \mathbf{f}_{n\mathbf{k}} \otimes \mathbf{f}_{n\mathbf{k}}^* . \quad \text{(D5)}$$

By substituting Eq. (D4) in Eq. (D3) it is found after some manipulations that the gap Chern number is $\mathcal{C} = \mathcal{C}_1 + \mathcal{C}_2$ with:

$$\mathcal{C}_1 = \frac{1}{2} \frac{1}{(2\pi)^2} \int d^2\mathbf{k} \int_{\omega_{gap}-i\infty}^{\omega_{gap}+i\infty} d\omega \left[ \left( \mathrm{tr}\left( \partial_1 \hat{N} \cdot \partial_\omega \overline{\mathbf{G}} \cdot \partial_2 \hat{N} \cdot \overline{\mathbf{G}} \right) - 1 \leftrightarrow 2 \right) \right.$$
$$\left. + \left( \frac{2k_1}{k^2 + k_{max}^2} \mathrm{tr}\left( \hat{N}(\mathbf{k}) \cdot \partial_\omega \overline{\mathbf{G}} \cdot \partial_2 \hat{N} \cdot \overline{\mathbf{G}} \right) - 1 \leftrightarrow 2 \right) \right. \quad \text{(D6a)}$$
$$\left. + \left( \frac{2k_2}{k^2 + k_{max}^2} \mathrm{tr}\left( \partial_1 \hat{N} \cdot \partial_\omega \overline{\mathbf{G}} \cdot \hat{N}(\mathbf{k}) \cdot \overline{\mathbf{G}} \right) - 1 \leftrightarrow 2 \right) \right]$$

$$C_2 = -\frac{1}{2}\frac{1}{(2\pi)^2}\int d^2\mathbf{k}\int_{\omega_{\text{gap}}-i\infty}^{\omega_{\text{gap}}+i\infty} d\omega\Big[$$

$$\frac{k_1}{k^2+k_{\max}^2}\Big(\operatorname{tr}\big(\mathbf{M}_\infty\cdot\partial_\omega[\omega\overline{\mathbf{G}}]\cdot\partial_2\hat{N}\cdot\overline{\mathbf{G}}\big)+\operatorname{tr}\big(\mathbf{M}_\infty\cdot(\partial_\omega\overline{\mathbf{G}})\cdot\partial_2\hat{N}\cdot(\omega\overline{\mathbf{G}})\big)+\operatorname{tr}\big((i\partial_\omega\overline{\mathbf{G}})\cdot\partial_2\hat{N}\big)-1\leftrightarrow 2\Big)$$

$$+\frac{k_2}{k^2+k_{\max}^2}\Big(\operatorname{tr}\big(\partial_1\hat{N}\cdot\partial_\omega[\omega\overline{\mathbf{G}}]\cdot\mathbf{M}_\infty\cdot\overline{\mathbf{G}}\big)+\operatorname{tr}\big(\partial_1\hat{N}\cdot(\partial_\omega\overline{\mathbf{G}})\cdot\mathbf{M}_\infty\cdot(\omega\overline{\mathbf{G}})\big)+\operatorname{tr}\big(\partial_1\hat{N}\cdot(i\partial_\omega\overline{\mathbf{G}})\big)-1\leftrightarrow 2\Big)\Big]$$

(D6b)

To obtain this formula we used the auxiliary identity in Eq. (F5) of Appendix F. Noting that $\int_{\omega_{\text{gap}}-i\infty}^{\omega_{\text{gap}}+i\infty} d\omega\,\partial_\omega\operatorname{tr}(\overline{\mathbf{G}}\cdot\partial_i\hat{N})=0$ and integrating by parts in frequency some of the terms of Eq. (D6b) one obtains Eq. (41b). Furthermore, integrating by parts the integral $\int_{\omega_{\text{gap}}-i\infty}^{\omega_{\text{gap}}+i\infty} d\omega\operatorname{tr}(\partial_i\hat{N}\cdot\partial_\omega\overline{\mathbf{G}}\cdot\partial_j\hat{N}\cdot\overline{\mathbf{G}})$ it is seen that it is anti-symmetric in the indices $i$ and $j$. Thus, using $\hat{N}(\mathbf{k})=k_1\partial_1\hat{N}+k_2\partial_2\hat{N}$ it is simple to verify that $C_1$ may be rewritten as in Eq. (41a). In Appendix F, we show that $\overline{\mathbf{G}}(\mathbf{k},\omega)$ may be explicitly calculated in terms of the nonlocal material matrix.

## Appendix E: Energy density flux

In this Appendix, we derive Eq. (D4). To begin with, we note that in the local case ($k_{\max}=\infty$) it follows from Eq. (D1) that $\frac{1}{2}\mathbf{Q}^*\cdot\partial_i\hat{L}_{\text{nl}}\cdot\mathbf{Q}=\frac{1}{2}\mathbf{f}^*\cdot\partial_i\hat{N}\cdot\mathbf{f}=\frac{1}{2}\hat{\mathbf{u}}_i\cdot(\mathbf{E}\times\mathbf{H}^*+\mathbf{E}^*\times\mathbf{H})$, and hence $\frac{1}{2}\mathbf{Q}^*\cdot\partial_i\hat{L}_{\text{nl}}\cdot\mathbf{Q}$ may be understood as the energy density flux (Poynting vector). Later, we show that this interpretation is still valid when $k_{\max}$ is finite. Crossed terms of the form $\mathbf{Q}_1^*\cdot\partial_i\hat{L}_{\text{nl}}\cdot\mathbf{Q}_2$ appear when one evaluates the Poynting vector associated with a linear combination of two fields. Here, we want to determine such crossed terms, $\mathbf{Q}_{n\mathbf{k}}^* \cdot \partial_i \hat{L}_{\mathrm{nl}} \cdot \mathbf{Q}_{m\mathbf{k}}$, with $\mathbf{Q}_{n\mathbf{k}}, \mathbf{Q}_{m\mathbf{k}}$ natural modes of $\hat{H}_{\mathrm{nl}}$.

To do this, we use $\mathbf{Q}_{n\mathbf{k}} = \begin{pmatrix} \mathbf{f}_{n\mathbf{k}} & \mathbf{Q}_{n\mathbf{k}}^{(1)} & \ldots \end{pmatrix}^T$ and Eq. (D1) to write:

$$\mathbf{Q}_{n\mathbf{k}}^* \cdot \partial_i \hat{L}_{\mathrm{nl}} \cdot \mathbf{Q}_{m\mathbf{k}} = \mathbf{f}_{n\mathbf{k}}^* \cdot \left[ \partial_i \hat{N} + \partial_i \left( \frac{1}{1 + k^2 / k_{\max}^2} \right) \sum_\alpha \mathrm{sgn}(\omega_{p,\alpha}) \mathbf{A}_\alpha^2 \right] \cdot \mathbf{f}_{m\mathbf{k}}$$
$$+ \mathbf{f}_{n\mathbf{k}}^* \cdot \sum_\alpha \partial_i \left( \frac{1}{\left(1 + k^2 / k_{\max}^2\right)^{1/2}} \right) |\omega_{p,\alpha}|^{1/2} \mathbf{A}_\alpha \cdot \mathbf{Q}_{m\mathbf{k}}^{(\alpha)} \quad \quad (\mathrm{E1})$$
$$+ \sum_\alpha \mathbf{Q}_{n\mathbf{k}}^{(\alpha),*} \cdot \partial_i \left( \frac{1}{\left(1 + k^2 / k_{\max}^2\right)^{1/2}} \right) |\omega_{p,\alpha}|^{1/2} \mathbf{A}_\alpha \cdot \mathbf{f}_{m\mathbf{k}}$$

From $\hat{H}_{\mathrm{nl}} \cdot \mathbf{Q}_{n\mathbf{k}} = \omega_{n\mathbf{k}} \mathbf{Q}_{n\mathbf{k}}$ and from the definition of $\hat{H}_{\mathrm{nl}}$ it is found that $\mathbf{Q}_{n\mathbf{k}}^{(\alpha)} = \frac{1}{\left(1 + k^2 / k_{\max}^2\right)^{1/2}} \frac{|\omega_{p,\alpha}|^{1/2}}{(\omega_{n\mathbf{k}} - \omega_{p,\alpha})} \mathbf{A}_\alpha \cdot \mathbf{f}_{n\mathbf{k}}$ (compare with Eq. (7)). Substituting this result into Eq. (E1) and using the Hermitian property of $\mathbf{A}_\alpha$, it is found after some simplifications that:

$$\mathbf{Q}_{n\mathbf{k}}^* \cdot \partial_i \hat{L}_{\mathrm{nl}} \cdot \mathbf{Q}_{m\mathbf{k}} = \mathbf{f}_{n\mathbf{k}}^* \cdot \partial_i \hat{N} \cdot \mathbf{f}_{m\mathbf{k}}$$
$$- \partial_i \left( \frac{1}{1 + k^2 / k_{\max}^2} \right) \left[ \frac{1}{2} \mathbf{f}_{n\mathbf{k}}^* \cdot \left(\mathbf{M}_{\mathrm{loc}}(\omega_{m\mathbf{k}}) - \mathbf{M}_\infty\right) \omega_{m\mathbf{k}} \cdot \mathbf{f}_{m\mathbf{k}} + \mathbf{f}_{n\mathbf{k}}^* \cdot \omega_{n\mathbf{k}} \frac{1}{2} \left(\mathbf{M}_{\mathrm{loc}}(\omega_{n\mathbf{k}}) - \mathbf{M}_\infty\right) \cdot \mathbf{f}_{m\mathbf{k}} \right]$$
(E2)

where $\mathbf{M}_{\mathrm{loc}}$ is defined as in Eq. (4), i.e., it corresponds to the material response in the $k_{\max} \to \infty$ limit [see Eq. (40)]. Using now Eq. (40), $\hat{N}(\mathbf{k}) \cdot \mathbf{f}_{n\mathbf{k}} = \omega_{n\mathbf{k}} \mathbf{M}(\mathbf{k}, \omega_{n\mathbf{k}}) \cdot \mathbf{f}_{n\mathbf{k}}$ and the fact that the relevant operators are Hermitian, we readily obtain the desired result [Eq. (D4)].

In the particular case $n = m$, Eq. (E2) is equivalent to $\frac{1}{2}\mathbf{Q}^*_{n\mathbf{k}} \cdot \partial_i \hat{L}_{nl} \cdot \mathbf{Q}_{n\mathbf{k}} = \frac{1}{2}\mathbf{f}^*_{n\mathbf{k}} \cdot \partial_i \hat{N} \cdot \mathbf{f}_{n\mathbf{k}} - \frac{1}{2}\omega_{n\mathbf{k}}\mathbf{f}^*_{n\mathbf{k}} \cdot \partial_i \mathbf{M}(\mathbf{k},\omega_{n\mathbf{k}}) \cdot \mathbf{f}_{n\mathbf{k}}$. The right-hand side of this expression gives precisely the Poynting vector (for a complex-valued field) in a generic spatially dispersive material [33, 34, 35, 36]. This confirms that $\frac{1}{2}\mathbf{Q}^* \cdot \partial_i \hat{L}_{nl} \cdot \mathbf{Q}$ may generally be understood as the Poynting vector.

## Appendix F: The 2D Green function in the nonlocal case

In what follows, we show that the Green function $\overline{\mathbf{G}}(\mathbf{k},\omega)$ defined as in Eq. (D5) satisfies:

$$\left[\hat{N}(\mathbf{k}) - \omega \mathbf{M}(\mathbf{k},\omega)\right] \cdot \overline{\mathbf{G}}(\mathbf{k},\omega) = i\mathbf{1}. \tag{F1}$$

The proof follows a sequence of steps analogous to Appendix A.

To begin with, first we note that the solution of $\left[\hat{N}(\mathbf{k}) - \omega \mathbf{M}(\mathbf{k},\omega)\right] \cdot \mathbf{f} = i\mathbf{j}$ (with $\mathbf{f}, \mathbf{j}$ constant vectors) can be found from the solution of the corresponding augmented problem $\hat{L}_{nl}(\mathbf{k}) \cdot \mathbf{Q} = \omega \mathbf{M}_g \cdot \mathbf{Q} + i\mathbf{j}_g$, with $\mathbf{j}_g$ defined as in Eq. (5) [11]. Furthermore, it was shown in Ref. [11, Ap. B] that if $\mathbf{f}_A$ and $\mathbf{f}_B$ satisfy $\left[\hat{N}(\mathbf{k}) - \omega \mathbf{M}(\mathbf{k},\omega_l)\right] \cdot \mathbf{f}_l = i\mathbf{j}_l$ ($l$=A,B) and if $\mathbf{Q}_A$ and $\mathbf{Q}_B$ are the corresponding solutions of the augmented problem, then:

$$\frac{1}{2}\mathbf{Q}^*_A \cdot \mathbf{M}_g \cdot \mathbf{Q}_B = \begin{cases} \frac{1}{2}\mathbf{f}^*_A \cdot \frac{\partial}{\partial \omega}\left[\omega \mathbf{M}(\mathbf{k},\omega)\right]_{\omega=\omega_A} \cdot \mathbf{f}_B, & \text{if } \omega_A = \omega_B \\ \frac{1}{2}\mathbf{f}^*_A \cdot \left[\frac{\omega_A \mathbf{M}(\mathbf{k},\omega_A) - \omega_B \mathbf{M}(\mathbf{k},\omega_B)}{\omega_A - \omega_B}\right] \cdot \mathbf{f}_B, & \text{if } \omega_A \neq \omega_B \end{cases}. \tag{F2}$$

As discussed in Appendix D, the operator $\hat{H}_{\mathrm{nl}} = \mathbf{M}_g^{-1} \cdot \hat{L}_{\mathrm{nl}}(\mathbf{k})$ is Hermitian with respect to the weighted inner product (6). For a continuum, it is more convenient to take the inner product as $\langle \mathbf{Q}_A | \mathbf{Q}_B \rangle_{\mathrm{co}} = \frac{1}{2} \mathbf{Q}_A^* \cdot \mathbf{M}_g \cdot \mathbf{Q}_B$. Let then $\{\mathbf{Q}_{n\mathbf{k}}\}_{n=1,2,\ldots}$ be a basis of the relevant (finite dimension) vector space normalized as $\langle \mathbf{Q}_{n\mathbf{k}} | \mathbf{Q}_{m\mathbf{k}} \rangle_{\mathrm{co}} = \delta_{m,n}$ and let $\mathbf{f}_{n\mathbf{k}}$ be the electromagnetic component of $\mathbf{Q}_{n\mathbf{k}}$. The completeness of the basis implies that $\mathbf{M}_g^{-1} = \frac{1}{2} \sum_n \mathbf{Q}_{n\mathbf{k}} \otimes \mathbf{Q}_{n\mathbf{k}}^*$. Hence, by projection, the electromagnetic modes satisfy the completeness relation:

$$\mathbf{M}_\infty^{-1} = \frac{1}{2} \sum_n \mathbf{f}_{n\mathbf{k}} \otimes \mathbf{f}_{n\mathbf{k}}^* . \tag{F3}$$

To find the solution of $\hat{L}_{\mathrm{nl}}(\mathbf{k}) \cdot \mathbf{Q} = \omega \mathbf{M}_g \cdot \mathbf{Q} + i \mathbf{j}_g$, the state vector is expanded into modes $\mathbf{Q} = \sum_n c_n \mathbf{Q}_{n\mathbf{k}}$. Clearly, $c_n = \frac{1}{2} \mathbf{Q}_{n\mathbf{k}}^* \cdot \mathbf{M}_g \cdot \mathbf{Q}$ which from Eq. (F2) may be written as:

$$c_n = \frac{1}{2} \mathbf{f}_{n\mathbf{k}}^* \cdot \left[ \frac{\omega_{n\mathbf{k}} \mathbf{M}(\mathbf{k}, \omega_{n\mathbf{k}}) - \omega \mathbf{M}(\mathbf{k}, \omega)}{\omega_{n\mathbf{k}} - \omega} \right] \cdot \mathbf{f} = \frac{1}{2} \mathbf{f}_{n\mathbf{k}}^* \cdot \frac{i\mathbf{j}}{\omega_{n\mathbf{k}} - \omega} . \tag{F4}$$

The second equality follows from $\left[ \hat{N}(\mathbf{k}) - \omega \mathbf{M}(\mathbf{k}, \omega) \right] \cdot \mathbf{f} = i\mathbf{j}$ and $\hat{N}(\mathbf{k}) \cdot \mathbf{f}_{n\mathbf{k}} = \omega_{n\mathbf{k}} \mathbf{M}(\mathbf{k}, \omega_{n\mathbf{k}}) \cdot \mathbf{f}_{n\mathbf{k}}$. By projecting $\mathbf{Q} = \sum_n c_n \mathbf{Q}_{n\mathbf{k}}$ into the electromagnetic subspace one finds $\mathbf{f} = \sum_n c_n \mathbf{f}_{n\mathbf{k}}$. Thus, from Eq. (F4) the solution of Eq. (F1) has indeed the modal expansion (D5), as we wanted to show.

Furthermore, using the completeness relation (F3) in Eq. (D5) it is found that:

$$i\omega \overline{\mathbf{G}}(\mathbf{k},\omega) - \mathbf{M}_\infty^{-1} = \frac{1}{2}\sum_n \frac{\omega_{n\mathbf{k}}}{\omega - \omega_{n\mathbf{k}}} \mathbf{f}_{n\mathbf{k}} \otimes \mathbf{f}_{n\mathbf{k}}^* .\tag{F5}$$

To conclude, we note that the solution of Eq. (F1) may be formally written as the inverse of a 6×6 matrix: $\overline{\mathbf{G}}(\mathbf{k},\omega) = i\left[\hat{N}(\mathbf{k}) - \omega \mathbf{M}(\mathbf{k},\omega)\right]^{-1}$. If we restrict ourselves to the sub-space formed by TM-polarized waves (so that the summation in Eq. (D5) only includes TM-polarized modes), then $\overline{\mathbf{G}}(\mathbf{k},\omega) = i\left[\hat{N}(\mathbf{k}) - \omega \mathbf{M}(\mathbf{k},\omega)\right]^{-1} \cdot \mathbf{1}_{\text{TM}}$. Note that $\mathbf{1}_{\text{TM}}$ may be regarded as a projection operator into the subspace of TM-waves.

-46-

[31] H. B. G. Casimir, and D. Polder, The Influence of Retardation on the London-van der Waals Forces, Phys. Rev. **73**, 360, (1948).

[32] M. G. Silveirinha, Casimir interaction between metaldielectric metamaterial slabs: Attraction at all macroscopic distances, *Phys. Rev. B*, **82**, 085101, (2010).

[33] L. D. Landau, E. M. Lifshitz, *Electrodynamics of Continuous Media,* Course of Theoretical Physics, vol.8, Elsevier Butterworth-Heinemann, 2004.

[34] V. Agranovich and V. Ginzburg, *Spatial Dispersion in Crystal Optics and the Theory of Excitons*. New York: Wiley- Interscience, 1966.

[35] M. G. Silveirinha, "Poynting vector, heating rate, and stored energy in structured materials: A first-principles derivation", *Phys. Rev. B*, **80**, 235120, (2009).

[36] J. T. Costa, M. G. Silveirinha, A. Alù, "Poynting vector in negative-index metamaterials", *Phys. Rev. B*, **83**, 165120 (2011).